\documentclass[10pt,aps,pra,twocolumn,amsmath,amssymb,superscriptaddress,showpacs,longbibliography]{revtex4-1}

\usepackage{hyperref}

\usepackage{subfigure}                      

\usepackage{amsmath,amssymb,amsfonts}       
\usepackage{graphicx}                       
\usepackage{siunitx}
\usepackage[english]{babel}

\begin{document}

\title{Optical--nanofiber--based interface for single molecules}

\author{Sarah M. Skoff}
\email {sarah.skoff@tuwien.ac.at}
\affiliation{Vienna Center for Quantum Science and Technology, Institute of Atomic and Subatomic Physics, Vienna University of Technology, Stadionallee 2, A-1020 Vienna, Austria}
\author{David Papencordt}
\affiliation{Vienna Center for Quantum Science and Technology, Institute of Atomic and Subatomic Physics, Vienna University of Technology, Stadionallee 2, A-1020 Vienna, Austria}
\author{Hardy Schauffert}
\affiliation{Vienna Center for Quantum Science and Technology, Institute of Atomic and Subatomic Physics, Vienna University of Technology, Stadionallee 2, A-1020 Vienna, Austria}
\author{Bernhard C. Bayer}
\affiliation{Electron Microscopy Group, Faculty of Physics, University of Vienna, Boltzmanngasse 5, A-1090 Vienna, Austria}
\author{Arno Rauschenbeutel}
\email{arno.rauschenbeutel@tuwien.ac.at}
\affiliation{Vienna Center for Quantum Science and Technology, Institute of Atomic and Subatomic Physics, Vienna University of Technology, Stadionallee 2, A-1020 Vienna, Austria}

\begin{abstract}
Optical interfaces for quantum emitters are a prerequisite for implementing quantum networks. Here, we couple single molecules to the guided modes of an optical nanofiber. The molecules are embedded within a crystal that provides photostability and, due to the inhomogeneous broadening, a means to spectrally address single molecules. Single molecules are excited and detected solely via the nanofiber interface without the requirement of additional optical access. In this way, we realize a fully fiber--integrated system that is scalable and may become a versatile constituent for quantum hybrid systems.
\end{abstract}
\pacs{42.50.-p, 42.81.Qb, 33.80.-b, 78.67.Bf}
\maketitle

\section{Introduction}
In recent years, single molecules in solids \cite{siyushev_molecular_2014, basche_direct_1995, basche_photon_1992,celebrano_single-molecule_2011,gerhardt_strong_2007,gerhardt_coherent_2010,kozankiewicz_single-molecule_2014} and other solid state quantum emitters such as color centers in diamond \cite{jelezko_read-out_2004,maurer_room-temperature_2012,neumann_multipartite_2008,pingault_all-optical_2014} and quantum dots \cite{hansom_environment-assisted_2014,delteil_generation_2016,Yalla_cavity_2014, Arcari_photonic_2014, Javadi_waveguide_2015} have gained increasing interest as building blocks for quantum networks \cite{nemoto_photonic_2014,kimble_quantum_2008}, quantum metrology \cite{doherty_electronic_2014,goldstein_environment-assisted_2011,acosta_electromagnetically_2013} and nanosensors \cite{faez_optical_2014, mazzamuto_single-molecule_2014, kucsko_nanometre-scale_2013}. For all these applications a strong light--matter interaction is essential. This can be achieved by coupling to a large ensemble of quantum emitters \cite{lukin_controlling_2001,reim_towards_2010}, by employing a cavity \cite{volz_nonlinear_2014,thompson_coupling_2013,ritter_elementary_2012,schietinger_one-by-one_2008}  or by decreasing the mode area of the interacting light field \cite{faez_coherent_2014,hwang_dye_2011,yalla_efficient_2012,vetsch_optical_2010, kirsanske_qd_2017,lombardi_molecules_2017} and hence achieving a significant overlap between the absorption cross--section of the emitter and the respective light field. A versatile platform to achieve such a small mode area of the light field are optical nanofibers \cite{vetsch_optical_2010,garcia-fernandez_optical_2011}. An optical nanofiber is the waist of a tapered optical fiber (TOF) and has a diameter smaller than the wavelength of the light it is guiding. Therefore, an appreciable fraction of the light propagates outside the fiber in the form of an evanescent wave. Due to the strong transverse confinement of the light field, which prevails over the entire length of the nanofiber, the interaction with emitters close to the surface can be significant \cite{le_kien_spontaneous_2005,reitz_coherence_2013,nayak_optical_2007,yalla_efficient_2012,liebermeister_tapered_2014}.

Single molecules in crystalline solids are efficient quantum emitters that exhibit strong zero phonon lines (ZPL) which can be lifetime-limited and as narrow as tens of MHz at cryogenic temperatures \cite{kummer_terrylene_1994, moerner_single-photon_2004}. Due to inhomogeneous broadening caused by the host crystal such molecules can be spectrally discerned and individually addressed using a narrowband laser \cite{moerner_illuminating_1999}. For a low concentration of molecules, this makes it possible to circumvent the additional spatial selection that has been used for numerous single molecule experiments in the past \cite{michaelis_single_1999,tamarat_ten_2000}, even if a large fraction of the crystal is illuminated. This was also exploited in recent experiments \cite{faez_coherent_2014, nanoguide_2017}, where single dibenzoterrylene molecules have been coupled to light propagating through a nanocapillary.
\begin{figure}[h]
\center
\includegraphics[width=\columnwidth]{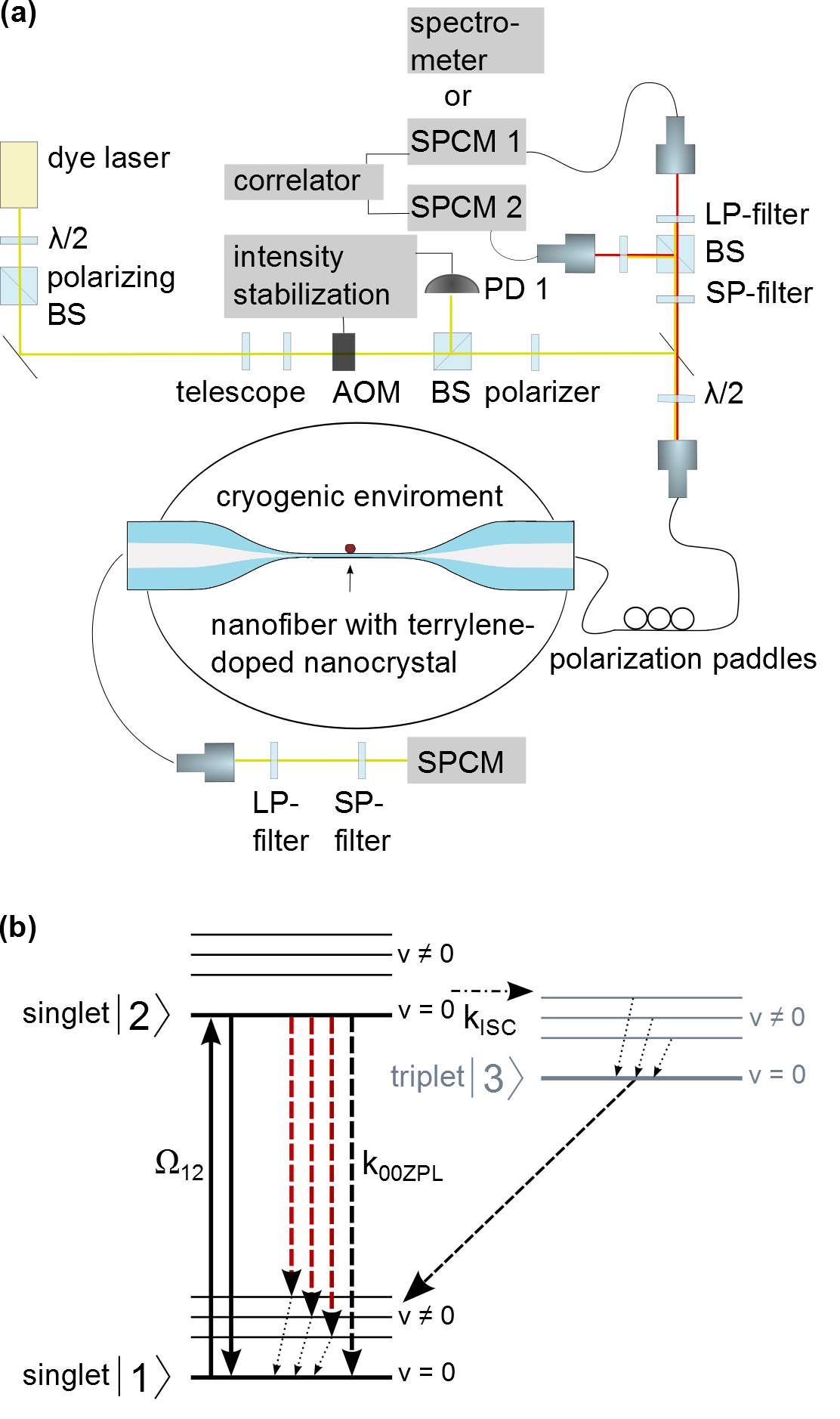}
\caption{(a) Experimental set--up for single molecule spectroscopy via an optical nanofiber interface. (b) Simplified level diagram of terrylene in p--terphenyl. Full lines indicate transitions driven by the laser, while dashed lines represent spontaneous emission. The dot--dashed line marked k$_{\text{ISC}}$ shows intersystem crossing from the excited singlet state to the triplet state. Dotted lines within an electronic manifold correspond to non--radiative decay processes that occur on a timescale of ps.}
\label{fig:setup}
\end{figure}
Single molecules come in a large variety and they are small quantum emitters which is useful when coupling them to nano- and microcavities \cite{coherent_microcavity_2017, emission_microcavity_2009, excitation_microcavity_1997} and also offers the possibility to study collective phenomena of quantum emitters. Additionally, single molecules such as polycyclic aromatic hydrocarbons can be spectrally very stable and do not suffer from photobleaching when embedded in the right host matrix. In addition to a near-unity quantum yield, these are very important features when working with solid state emitters.
Here, we show for the first time that single organic molecules can be interfaced with an optical nanofiber. This presents a new platform based on solid state emitters that can be used for quantum optics and that is naturally integrated into optical fiber networks.

\section{Experimental Setup}
In our experimental setup, the TOF resides inside a cryostat [Fig. \ref{fig:setup}(a)] and we interface terrylene molecules in a para--terphenyl (p--terphenyl) crystal with the evanescent light field surrounding its nanofiber. The latter has a total length of \SI{3}{mm} and a diameter of \SI{320}{nm}. The TOF is produced in a heat and pull process using a custom--made pulling rig \cite{warken_fiber_2008}.
In the \SI{6.8}{cm} long tapered section of the fiber, the weakly guided LP$_{01}$ mode of the standard single mode optical fiber is adiabatically transformed into the strongly guided HE$_{11}$ mode of the nanofiber waist and back yielding transmission losses of less than 2\% from 520--\SI{650}{nm}. For our purpose, a broadband transmission is crucial as the excitation and detection wavelengths can differ by more than 100 nm. This requires a careful choice of tapering angles and waist diameter \cite{stiebeiner_design_2010}.
Terrylene in p--terphenyl can exhibit four different electronic transition frequencies from the ground to the first excited state termed X1--X4, corresponding to four possible orientations of the molecules in the crystal.  Molecules in the X4 orientation are resonant with light at \SI{577.9}{nm} and have been shown to be very photostable \cite{bordat_molecular_2002}. Hence, all measurements presented here use molecules in this site. A simplified level diagram is depicted in Fig. \ref{fig:setup}(b). The laser excites the molecule on the zero phonon line 00ZPL that connects the ground and the excited electronic state without any vibrational contribution of the molecule. After excitation, the molecule will decay into any of the vibrational states in the electronic ground state with a probability determined by the Franck Condon $\alpha_{\text{FC}}$ and Debye--Waller $\alpha_{\text{DW}}$ factors. From these states it will nonradiatively decay into the vibrational ground state within picoseconds. Hence, the absorption cross--section \cite{loudon_quantum_2000} of a single molecule in a solid is
\begin{equation}
\sigma = \alpha_{\text{FC}} \,\alpha_{\text{DW}} \frac{3 \lambda^2}{2 \pi} \frac{\Gamma^2}{\Gamma_{\text{hom}}^2} (\mathbf{\hat{d}}\cdot \mathbf{\hat{e}})^2
\end{equation}
where the product $\mathbf{\hat{d}}\cdot \mathbf{\hat{e}}$ represents the projection of the polarization vector $\mathbf{\hat{e}}$ of the excitation light on the unit vector of the molecular dipole $\mathbf{\hat{d}}$ and $\lambda$ denotes the excitation wavelength. $\Gamma$ is the lifetime--limited linewidth and $\Gamma_{\text{hom}}$ the homogeneously broadened linewidth of the 00ZPL, respectively.
If the molecular dipole is aligned with the polarization of the excitation light and the homogeneous broadening is negligible, the absorption cross--section will approach that of a simple two--level atom and is comparable to the effective mode area $A_{\text{eff}}$ \cite{warken_ultra-sensitive_2007} of our optical nanofiber of about $0.4 \times \lambda^{2}$ (see Appendix \ref{app:excitation}). This ensures a strong effect of a single molecule on the light field.
The probability of a terrylene molecule to decay to the triplet state after excitation rather than to the singlet ground state is very low and has experimentally been found to be $< 10^{-5}$ at cryogenic temperatures \cite{banasiewicz_triplet_2005,basche_direct_1995}. 
\begin{figure}[h!]
\center
\includegraphics[width=\columnwidth]{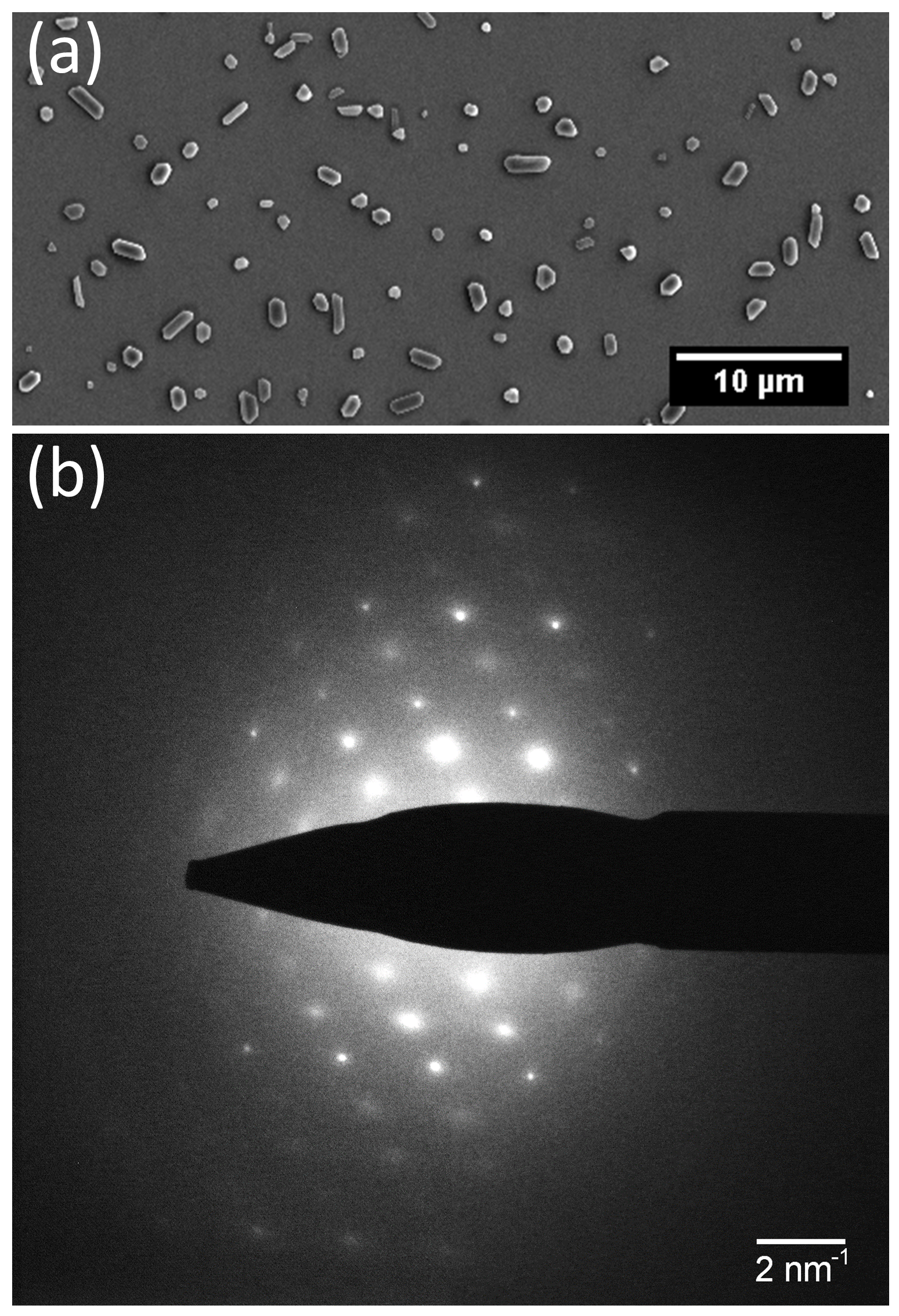}
\caption{(a) Scanning electron microscope image (FEI Quanta 200 FEG) of terrylene--doped p--terphenyl crystals on a silicon substrate. (b) Selected area electron diffraction pattern of an individual terrylene--doped p--terphenyl crystal (Philips CM200 TEM at 200 kV, imaged with the electron beam perpendicular to the crystal base face, exposure time was kept short to avoid electron beam induced degradation of the organic crystal \cite{li_radiation_2004}). The pattern is consistent with the monoclinic high--temperature phase of a single p--terphenyl single--crystal \cite{rice_temperature_2012} viewed along the [001] axis (considering double diffraction effects, Cambridge Structural Database (CSD) entry terphe14).}
\label{fig:crystals}
\end{figure}
To maintain sufficient light guiding capabilities of the nanofiber--crystal system for spectroscopy and to ensure that the crystal stays tightly adhered to the vertically mounted nanofiber, the crystals have to be on the order of a few hundred nanometers in size. Such nanocrystals are grown by a reprecipitation method \cite{kasai_novel_1992} from an oversaturated solution of a $8 \times 10^{-5}$ molar mixture of terrylene/p--terphenyl in toluene. The solution is heated until both compounds are dissolved and then isopropanol is added as a reprecipitation agent. This procedure results in terrylene doped p--terphenyl crystals of platelet morphology as seen in Fig. \ref{fig:crystals}(a), which shows a scanning electron microscope (SEM) image of such crystals deposited on a silicon substrate. The majority of crystals have base dimensions in the range of 200 to 2000 nm and a base width to height ratio of 2.5:1 to 5:1 as determined by atomic force microscopy (AFM) measurements. The single-crystalline nature of these crystals has been verified by performing selected area electron diffraction (SAED) measurements using a transmission electron microscope (TEM), see Fig. \ref{fig:crystals}(b). SAED also indicates that the substrate--supported crystal platelet base face is of (001) orientation, i.e. the crystal's c-axis is perpendicular to the substrate. It is known that the dipole moment of the transition to the lowest electronically excited state in terrylene is linear and lies along the long axis of the molecule \cite{bordat_molecular_2002}. When inserted into a p--terphenyl crystal, the molecule's long axis and thus the dipole moment lie nearly parallel to the crystal's c-axis and therefore in our configuration nearly perpendicular to the substrate.

A terrylene--doped p--terphenyl nanocrystal is deposited on the nanofiber by a drop--touch method: a drop of the suspension of doped p--terphenyl nanocrystals is briefly brought into contact with the nanofiber via a pipette. During this process, the transmission of the excitation laser and the fluorescence of the nanofiber is monitored with a power meter and a spectrometer, respectively. When a doped crystal has adhered to the nanofiber surface during the contact with the suspension droplet, a typical fluorescence signal and some loss in transmission is observed. As crystals of a variety of sizes are produced during our growth process, we have to post--select the size of the deposited crystal. However, since the largest crystals sediment faster, we usually find suitably small ones in the suspension supernatant. If it is nevertheless found that the transmission deteriorates too much during the deposition process, the crystal can be washed off with acetone and another one deposited.

The TOF is mounted on a steel holder with two NdFeB magnets. This ensures that the fiber is firmly held and that it stays intact during the cooling process down to cryogenic temperatures. This fiber setup is mounted in the cold--pot of a custom--made cryostat that can cool the sample to \SI{4}{K}. To ensure efficient thermalisation of the nanofiber and the crystal with the walls of the cold--pot, helium buffer gas at a pressure of a few mbar is introduced into the cold--pot before cool--down, after it has been evacuated. 

To excite the molecules, light of a dye laser (Spectra-Physics Matisse-DS) is coupled into the optical fiber that connects to the TOF and that enters the cryostat via a teflon feed--through [Fig. \ref{fig:setup}(a)]. To compensate for intensity fluctuations and drifts, the laser beam is sent through an acousto--optic modulator (AOM) and partly onto a photodiode (PD) and is actively intensity stabilised. A fraction of the Stokes--shifted laser-induced fluorescence (LIF) of the molecule is collected by the nanofiber, fiber--guided out of the cryostat, and can be monitored by a spectrometer (Shamrock SR-303i, Andor Technology) or single photon counting modules (SPCMs) at either end of the optical fiber.
\begin{figure}[htb]
\center
\includegraphics[width=\columnwidth]{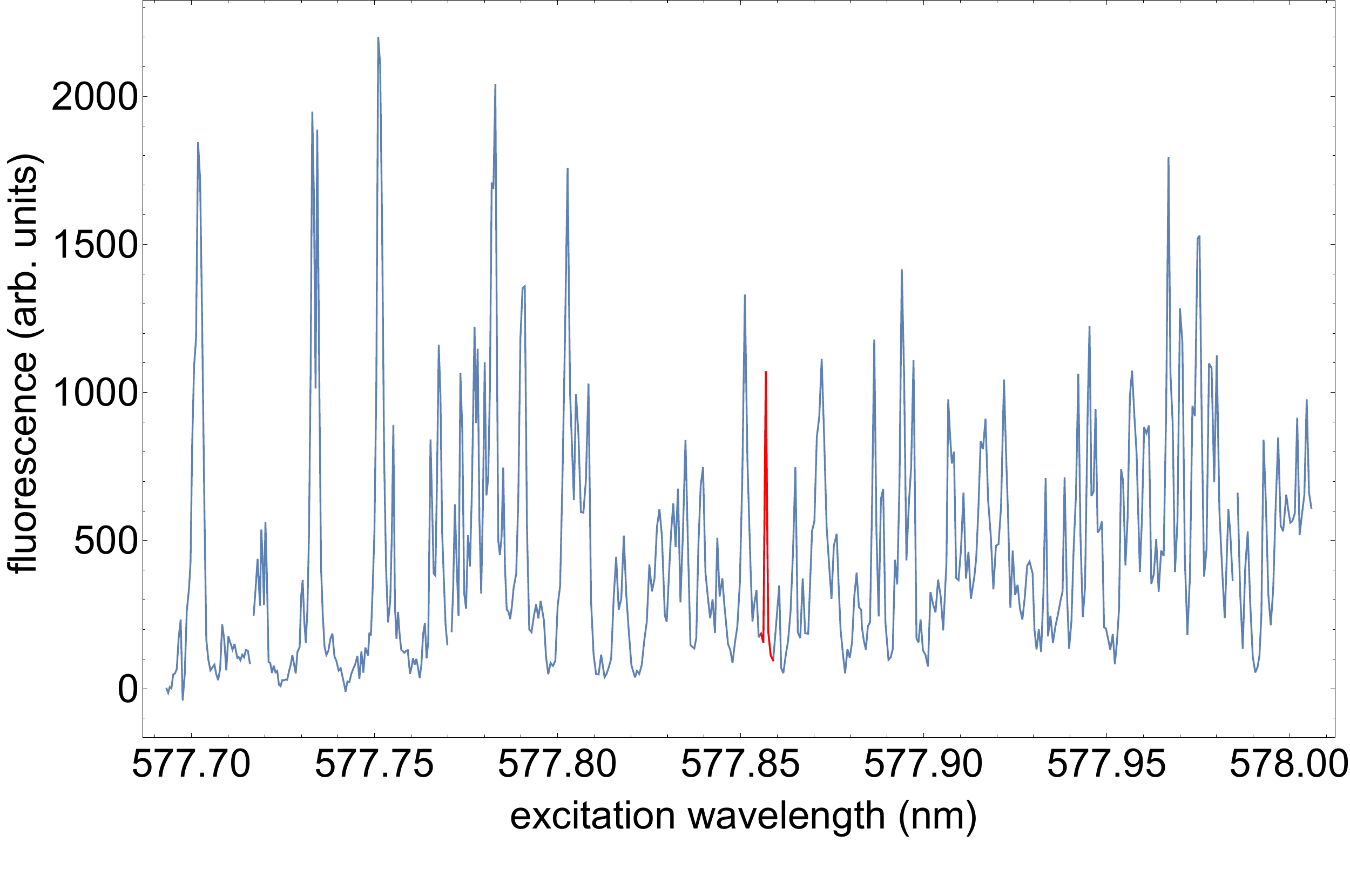}
\caption{Fluorescence excitation measurement of nanofiber--interfaced terrylene molecules. The red peak indicates the fluorescence of a single terrylene molecule.}
\label{fig:moleculesEnsemble}
\end{figure}
Contributions from the excitation light and fluorescence on the 00ZPL are filtered out by a long--pass (LP) filter. Together with a short--pass (SP) filter  to block Raman scattering from the fiber, this leaves a transmission window in the range of 630--650 nm. Due to the inhomogeneous line shifts induced by the host crystal matrix, single molecules can be spectrally selected with the narrowband dye laser if the terrylene concentration in the host crystal is small enough. 

\section{Experimental Results}
Figure \ref{fig:moleculesEnsemble} shows the fluorescence excitation spectrum of a molecular ensemble in the X4 orientation. The excitation line of what has been verified to be a single molecule is highlighted in red.

\begin{figure}[htb]
\center
\includegraphics[width=\columnwidth]{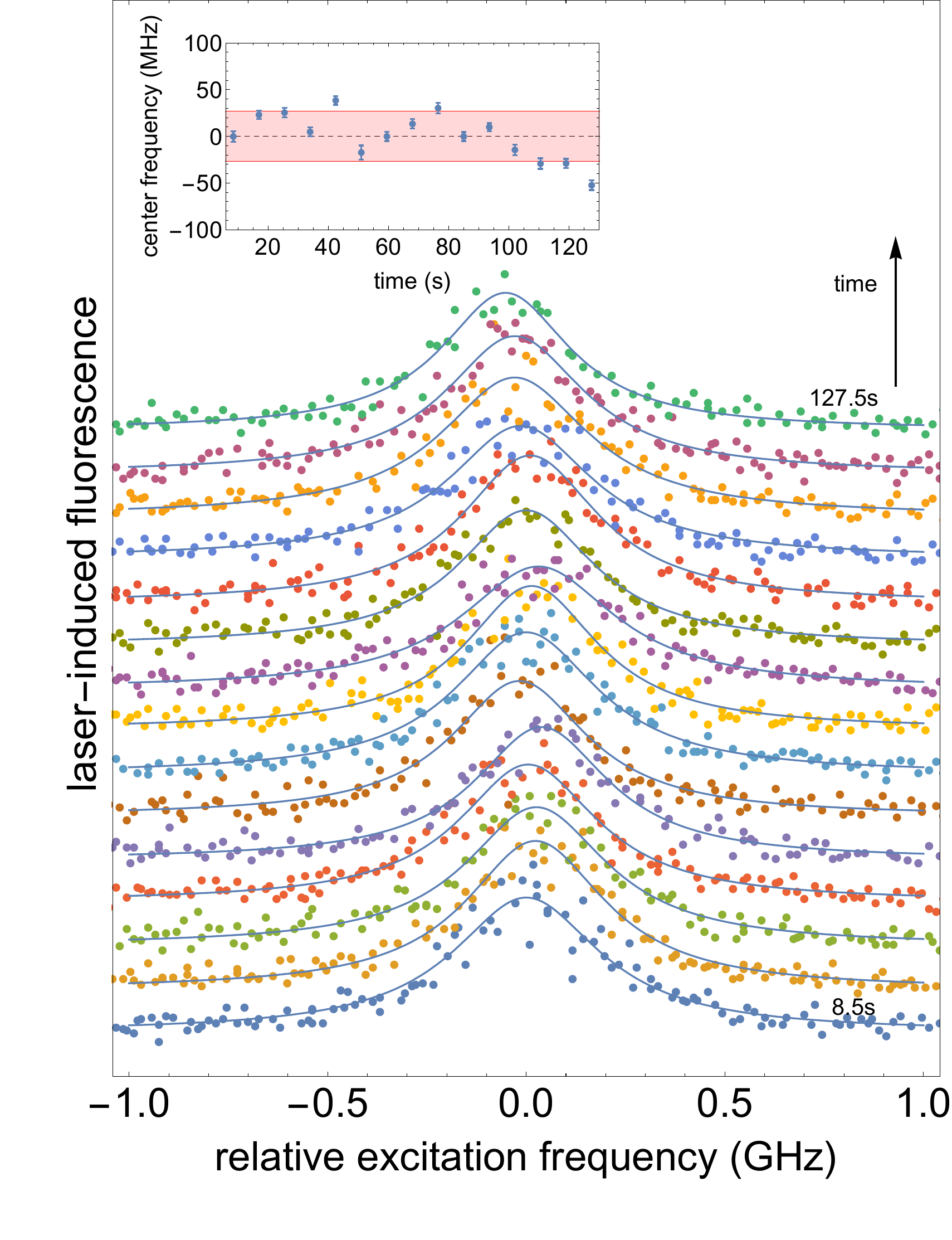}
\caption{Consecutive laser-induced fluorescence spectra of a single terrylene molecule coupled to an optical nanofiber. The individual scans are offset for clarity. Each scan takes 8.5 s. The inset explicitely shows the fitted resonance frequency as a function of time. The shaded region corresponds to the FWHM linewidth of a lifetime-limited molecular spectrum (see Appendix \ref{app:photons}).}
\label{fig:spectralDiffusion}
\end{figure}

To characterize the optical interface created by individual molecules and the optical nanofiber, we investigate several molecules that are located in the same nanocrystal. The transition frequencies of solid state quantum emitters are known to be very sensitive to the environment, which can be a favourable effect if controlled well \cite{dolde_electric-field_2011} or lead to unwanted spectral diffusion. In order to evaluate the stability of the transition frequency in our system, we record the same spectral line of the spectra of individual molecules over time. Figure \ref{fig:spectralDiffusion} shows the evolution of a spectral line over the timescale of minutes. The excitation power corresponds to a saturation parameter of $I/I_{\text{S}} = 0.7$ and, over all spectra, the linewidth was measured to be $429 \pm 20$ MHz. A slow drift of $0.44 \pm 0.14$ MHz/s is observable. 
This could be a result of a drift in the laser frequency, the wavemeter, or the molecular transition frequency. The faster frequency scatter on top of this drift is found to be $18$ MHz rms. This value is well below the lifetime-limited linewidth of the molecular transition (see Appendix \ref{app:photons}). Even for a lifetime-limited molecular spectrum and without any active stabilization this implies that the molecule would be resonant with a fixed-frequency laser for about two minutes. The measured stability of the molecule is thus found to be superior to other solid state emitters, which may require active stabilization of the resonance frequency on a faster timescale \cite{nvcenters_control_2010, qdot_stabilization_2014}. By cooling the doped crystal to 1.7 K, we can also achieve stable lifetime-limited spectra of our single molecules recorded on either end of the TOF, as is experimentally shown in Appendix \ref{app:photons}. 

\begin{figure}[htb]
\center
\includegraphics[width=\columnwidth]{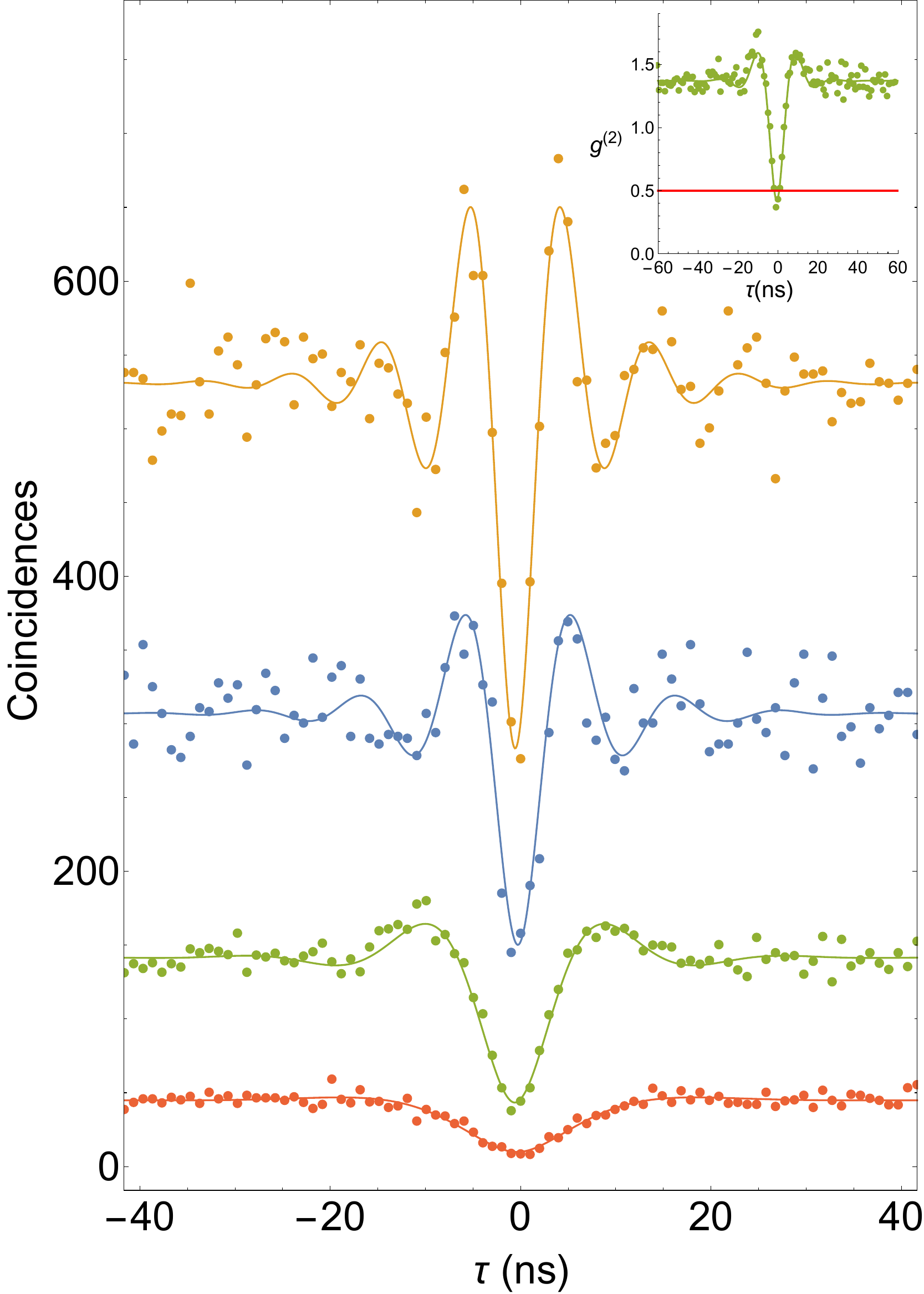}
\caption{Fluorescence intensity correlation measurements of molecule C for different excitation powers (from bottom to top: 0.7 nW, 2.9 nW, 7.4 nW, 11.1 nW). The coincidences were recorded for 1000 s and the time resolution for the data is 1 ns. With increasing excitation power, the onset of Rabi oscillations and an increase in incoherent background scatter is clearly seen. The inset shows the $g^{(2)}$-measurement at 2.9 nW normalised to the steady--state correlations of the fluorescence of the molecule, clearly indicating a single emitter.}
\label{fig:autocorrelation}
\end{figure}

To verify the interaction with single quantum emitters, second order fluorescence intensity correlation measurements are performed using a Hanbury-Brown-Twiss setup [Fig. \ref{fig:setup}(a)]. The backscattered fluorescence is split by a 50/50 beamsplitter (BS) and directed to two SPCMs. Their counts are then recorded by means of a field programmable gate array and correlated. For a single quantum emitter, photon antibunching is observed, which manifests itself as an antibunching dip in the intensity correlation measurements at zero time delay $\tau$ between two detection events. Experimental imperfections such as incoherent background scatter will reduce the contrast of the dip but a dip deeper than 50\% is prove of a single quantum emitter. 
Figure \ref{fig:autocorrelation} shows second order fluorescence intensity correlation measurements of a molecule on the nanofiber where the antibunching dip is clearly visible. 

\begin{figure}[h]
\center
\includegraphics[width=\columnwidth]{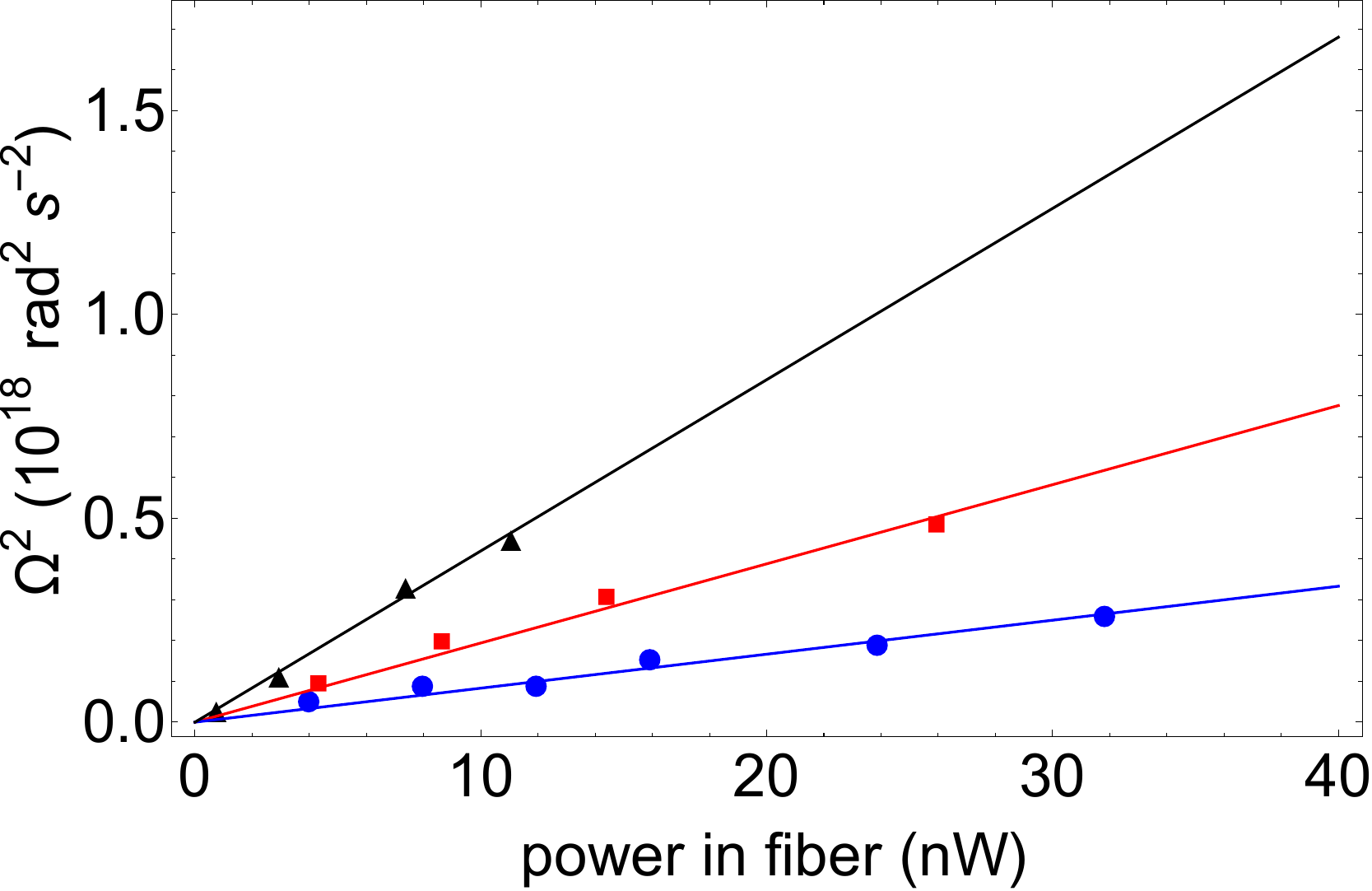}
\caption{Squared Rabi frequency as a function of excitation power for molecule A (red, squares), B (blue, circles) and C (black, triangles) and corresponding linear fits.}
\label{fig:RabiFrequency}
\end{figure}

\begin{figure}[h]
\center
\includegraphics[width=\columnwidth]{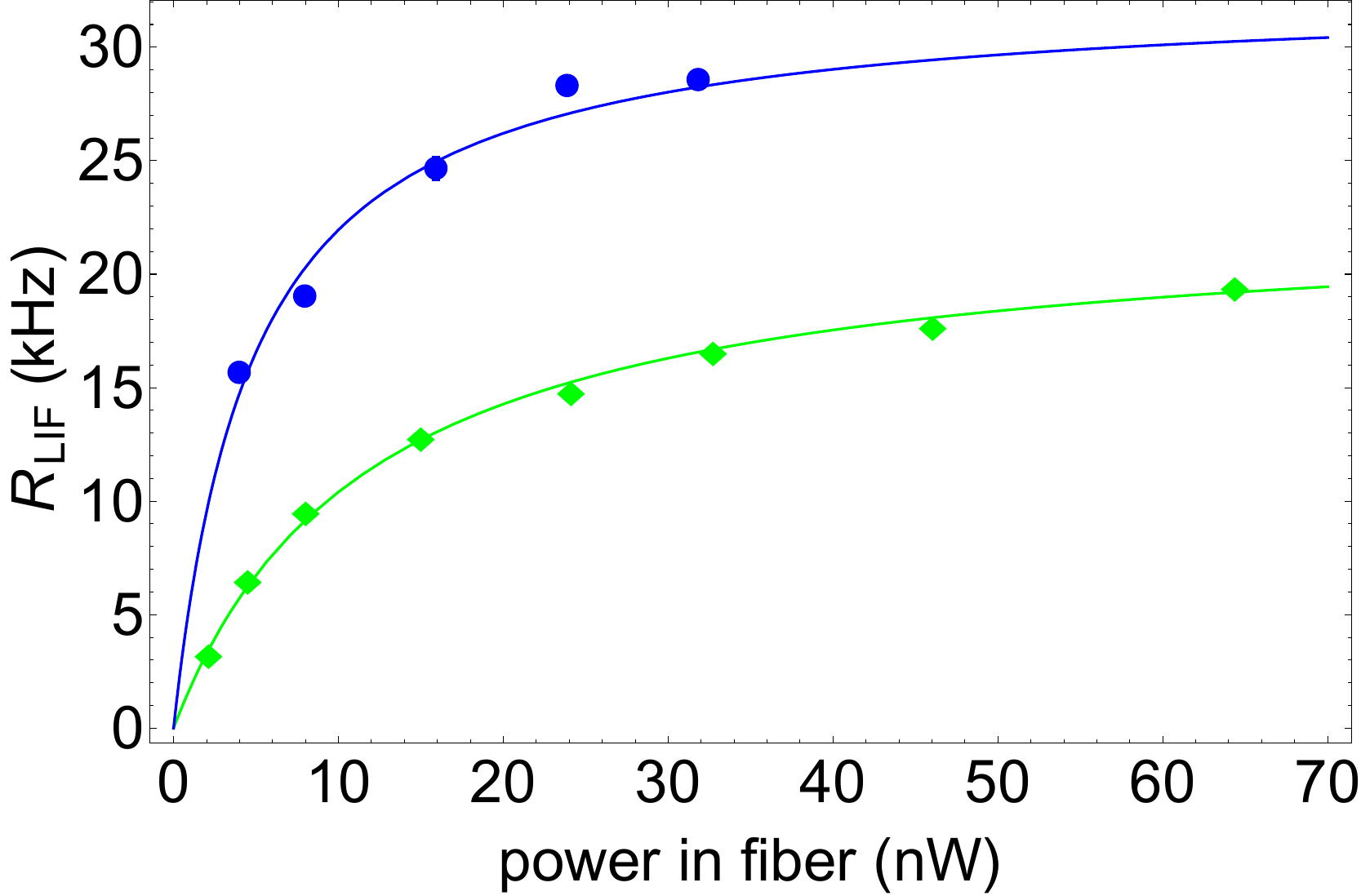}
\caption{Saturation of the resonant fluorescence intensity $R_{\text{LIF}}$ as the excitation laser power is increased for molecules B (blue, circles) and D (green, diamonds) and corresponding fits.}
\label{fig:saturation}
\end{figure}

As the excitation intensity is increased, the onset of Rabi oscillations is apparent. Since the decay into the triplet state is negligible for short times, the measurements are fitted to the correlation function that is obtained by solving the optical Bloch equations for a two level system with an overall amplitude $A$, a spontaneous decay rate $\Gamma$ and a Rabi frequency $\Omega = \mathbf{d}\cdot \mathbf{E}/\hbar$, where $\mathbf{d}$ is the dipole moment of the molecule's transition and $\textbf{E}$ the electric field at the molecule's location \cite{loudon_quantum_2000}.
\begin{equation}
\begin{aligned}
&\ g^{(2)} (\tau) =  A \Big(1-\Big[\cos(\nu \tau) + \frac{3 \Gamma}{{4}\nu}\sin(\nu \tau)\Big]\exp{\Big(\frac{-3 \Gamma \tau}{4}\Big)}\Big)\\
&\ \textnormal{with}\\
&\ \nu = \sqrt{\Omega^{2}-\Big(\frac{\Gamma}{4}\Big)^{2}}
\end{aligned}
\label{eq:gtwofunction}
\end{equation}
An incoherent Poissonian background $B$ added to a signal with average intensity $\langle I \rangle$ reduces the contrast of the intensity correlation measurements. This is taken into account when analyzing our data with the modified intensity correlation function \cite{verberk_photon_2003}
\begin{equation}
g^{(2)}_{B}(\tau) = 1+\frac{\langle I\rangle^2}{(\langle I \rangle+B)^2}(g^{2}(\tau)-1).
\label{eq:gtwofunctionB}
\end{equation} 
We fit eqn.(\ref{eq:gtwofunctionB}) to the data for different excitation powers while using $\Gamma$ as a global fit parameter for all measurements on a given molecule. This yields the respective Rabi frequencies and the non--power broadenend homogeneous linewidth $\Gamma$ of the molecule. From these fits we also obtain the saturation intensity $I_\text{S}$ of the molecule as $I/I_\text{S} = 2 \Omega^2/\Gamma^2$.
Figure \ref{fig:RabiFrequency} shows this expected linear increase of the squared Rabi frequency as a function of excitation power for three different molecules in the same crystal. The error bars obtained from fitting the intensity correlation measurements are smaller than the depicted datapoints in Fig. \ref{fig:RabiFrequency}. From the fits, the saturation power $P_{\text{S}}$ corresponding to $I_{\text{S}}$ is obtained. We obtain saturation powers for molecule A of ${P_\text{S}} = 1.8\pm 0.3$ nW, molecule B of ${P_\text{S}} = 4.3 \pm 0.8$ nW and molecule C of ${P_\text{S}} =0.5 \pm 0.1$ nW. The biggest contribution to the error in the saturation powers arises from the uncertainty in determining the power in the fiber. We convert $P_{\text{S}}$ to the maximum intensity at the surface of the nanofiber by considering the fundamental quasilinearly polarised HE$_{11}$ mode that is supported by our optical nanofiber with a diameter of 320 nm \cite{le_kien_spontaneous_2005}. Without the exact knowledge of the orientation of the molecule's transition dipole moment and its distance from the nanofiber surface, this gives an upper limit for the saturation intensity. The measured saturation intensities are $I_{\text{S}}<1.8$ Wcm$^{-2}$ for molecule A, $I_{\text{S}}<4.3$ Wcm$^{-2}$ for molecule B and $I_{\text{S}}<0.5$ Wcm$^{-2}$ for molecule C. These results compare well with results obtained by other groups who studied terrylene in bulk p--terphenyl \cite{kummer_terrylene_1994,tamarat_pump-probe_1995}. To our knowledge a saturation intensity of $I_{\text{S}}<0.5$ Wcm$^{-2}$ as for molecule C is the lowest measured so far for terrylene in p--terphenyl. As opposed to measurements on terrylene in bulk p--terphenyl using a confocal microscope, the excitation light in our case enters through the side of the thin host crystal platelets. As the transition dipole moment of the molecules lies nearly perpendicular to the base of these platelets, this suggests an improved overlap between the polarization of the nanofiber--guided excitation light and the transition dipole moment of the molecules.
An independent measurement of the saturation power is obtained by recording the resonant fluorescence rate $R_{\text{LIF}}$ as a function of excitation power as plotted in Fig. \ref{fig:saturation}, which includes a fit to $R_{\text{LIF}} = R_\infty \frac{P/P_\text{S}}{1+P/P_\text{S}}$. The error bars on the fluorescence rate that are obtained by taking the standard error of the amplitude from fits over several molecular spectra are smaller than the depicted datapoints. This measurement yields a saturation power of $4.8 \pm 1.4 $ nW for molecule B, where the error stems from the fit and from the uncertainty in the excitation power inside the fiber. This translates into a saturation intensity of $I_{\text{S}}<4.8$ Wcm$^{-2}$ for molecule B in good agreement with $I_\text{S}<4.3$ Wcm$^{-2}$ as obtained with the HBT setup. 
We performed a further measurement on a fourth molecule (molecule D) that yielded a lower fluorescence rate and indeed was measured to have a higher saturation power of $11.4 \pm 2.2 $ nW and therefore a saturation intensity of $<11.4$ Wcm$^{-2}$. This suggests that this molecule is located further away from the nanofiber surface such that its fluorescence does not couple back to the nanofiber--guided modes as efficiently.  Alternatively, the alignment between polarization of the excitation light and the dipole moment of molecule D may be less favorable. Although all molecules are embedded in a single crystal and therefore have the same orientation, the inherent birefringence of the host crystal can cause a less favorable alignment: The phase shift between the corresponding polarization components is $\Delta \phi = 2 \pi \Delta n L/\lambda$, where $\lambda$ is the vacuum wavelength, $L$ the propagation distance and $\Delta n$ is the effective birefringence that can reach 0.32 in our case.
\begin{figure}[htb]
\center
\includegraphics[width=\columnwidth]{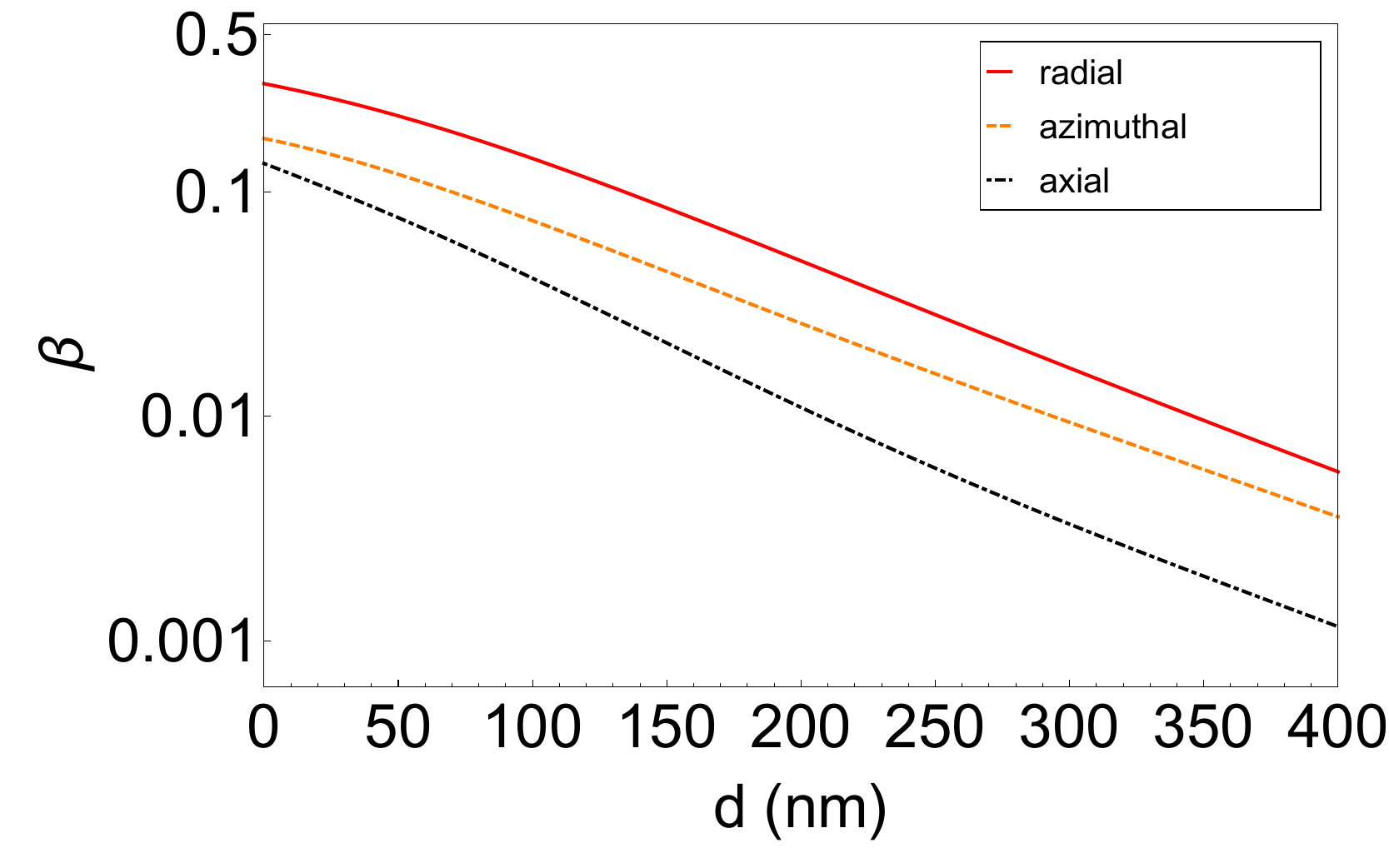}
\caption{Mean coupling efficiency of a linear radial (red), tangential (orange, dashed) and axial (black, dot--dashed) dipole, radiating at 630--650 nm, to the nanofiber as a function of distance from the nanofiber surface.}
\label{fig:dipoleNanofiber}
\end{figure}
The efficiency of exciting the different molecules is given by $\eta_{\text{abs}} = \sigma/A_{\text{eff}}(x,y) $, where $A_{\text{eff}}(x,y)$ is the effective mode area at the position of the molecule. More details can be found in Appendix \ref{app:excitation}. The molecules are also detected via the nanofiber interface and hence the overall efficiency for fluorescence excitation and detection via the nanofiber interface is then given as $\eta_{\text{LIF}} = \eta_{\text{abs}}\, \beta$, where $\beta = \Gamma_g/\Gamma_{sc}$ is the coupling efficiency of dipole radiation to the nanofiber modes. This coupling efficiency depends on the radiated wavelength, distance and orientation of the dipole with respect to the nanofiber surface. Here, $\Gamma_g$ is the scattering rate into guided modes and $\Gamma_{sc}$ is the total scattering rate of the dipole. For our case of a nanofiber with \SI{160}{nm} radius and Stokes shifted fluorescence in the range of 630--650 nm, we calculated this coupling efficiency for a radially, azimuthally and axially oriented dipole following \cite{le_kien_spontaneous_2005}, see Fig. \ref{fig:dipoleNanofiber}. 

\begin{figure}[htb]
\center
\includegraphics[width=\columnwidth]{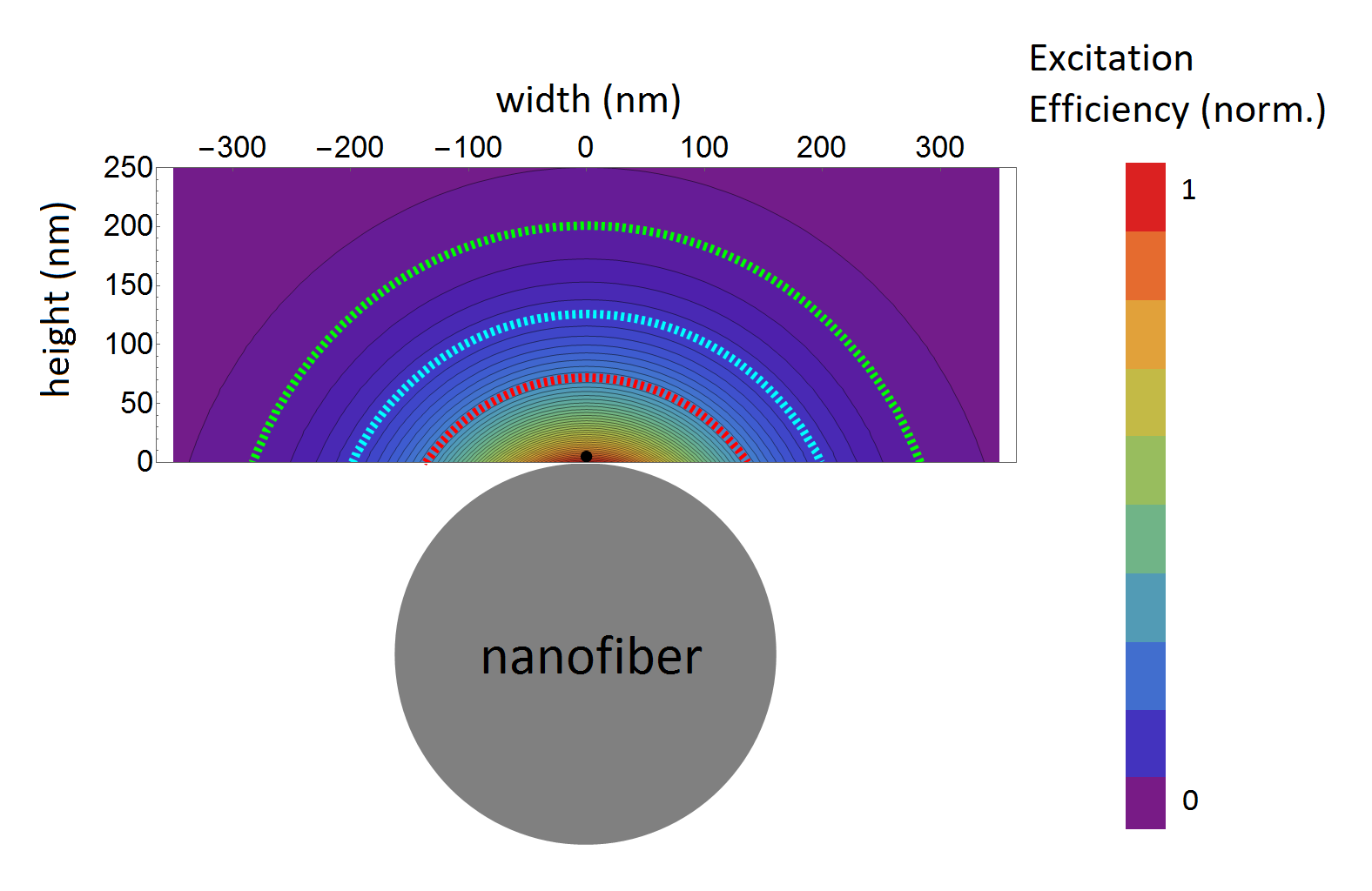}
\caption{The excitation efficiency $\eta_{\text{abs}}$ for a molecular dipole at different positions with respect to the nanofiber. The spatial coordinates are chosen assuming a platelet crystal that is lying with its base on the nanofiber surface.
The contact point of nanofiber and crystal is at the origin of this reference frame. The efficiency is normalised to the maxmimum excitation efficiency on the nanofiber surface of 17\% (see Appendix \ref{app:excitation}).  The maximum radial distances for the different molecules are indicated by dashed colored lines (A = red, B = blue, D = green, from bottom to top). The reference molecule C is shown as a black dot and is assumed to be located directly on the nanofiber surface at position (x,y) = (0,0).}
\label{fig:moleculeLocation}
\end{figure}

Since the power needed to saturate molecule C is the lowest yet measured for terrylene in p--terphenyl, we assume that this molecule is very close to the surface of the nanofiber. An upper limit for the other molecules from the nanofiber surface can then be estimated by comparing their saturation intensities. Because the host crystal is birefringent, we only give an upper limit on the radial distance between the different molecules. Figure \ref{fig:moleculeLocation} shows the calculated efficiency of fluorescence excitation via the nanofiber interface for different positions of the dipole with respect to the nanofiber surface. 
The positions are chosen to lie within the volume of a platelet crystal with its base on the nanofiber surface. The molecular dipole is oriented perpendicular to the crystal's base and excited by quasilinearly polarised light via the nanofiber--based interface. Assuming unpolarised light instead of quasilinearly polarised light affects the relative distances between the different molecules by less than 3\%.
We did not incorporate the refractive indices of the crystal into this model because they would make the local efficiencies very dependent on the crystal's geometry and we are only interested in assigning the maximum radial distance of the measured molecules. The maximum radial distances of the investigated molecules are depicted in their respective colours by dashed contours [Fig. \ref{fig:moleculeLocation}]. These results show that they radially all lie within less than \SI{481}{nm} of each other and less than \SI{201}{nm} from the nanofiber surface. This translates into coupling efficiencies for a radial dipole to the nanofiber mode between 5-30\% [Fig. \ref{fig:dipoleNanofiber}]. This means that our set-up can be a superior choice for coupling single photons to single mode optical fibers compared to using conventional confocal microscopes \cite{fujiwara_highly_2011, solid_immersion_2011, karlsson_confocal_2016} and thus opens the way for fully fiber--coupled single photon sources. It is also an important step towards strong coupling of single molecules to optical waveguide structures.

\section{Conclusion}   
Summarizing, we have shown how single molecules can be optically interfaced via the evanescent field surrounding an optical nanofiber. This is an important addition to the toolbox of quantum emitters such as atoms \cite{vetsch_optical_2010,goban_demonstration_2012,hafezi_atomic_2012}, quantum dots \cite{yalla_efficient_2012,fujiwara_highly_2011} and NV centers in nanodiamonds \cite{liebermeister_tapered_2014,fujiwara_ultrathin_2015} that have been fiber--integrated by coupling to optical nanofibers. Each of these systems has its own intrinsic advantages for their usage in quantum networks. Single molecules in solids are efficient quantum emitters that come in a large variety of emission wavelengths. This makes them suitable to be interfaced with other quantum emitters \cite{siyushev_molecular_2014}. They have an advantageous level structure for the implementation of triggered single photon sources \cite{moerner_single-photon_2004,lounis_single_2000} and have proven their versatility in quantum optics \cite{hwang_single-molecule_2009,pototschnig_controlling_2011,maser_few-photon_2016}. Further, single waveguide--coupled molecules allow the investigation of photon--mediated interactions between two quantum emitters even when they are separated by much more than the excitation wavelength \cite{rist_photon-mediated_2008,loo_photon-mediated_2013,le_kien_nanofiber-mediated_2005}. These interactions can be further enhanced by using a nanofiber between two fiber Bragg gratings and thereby realizing a high--Q cavity \cite{wuttke_nanofiber_2012}.
Single molecules that are coupled to the evanescent field of optical nanofibers therefore not only offer a rich experimental platform for investigating entanglement and correlations between quantum emitters, they also provide a means for implementing components of quantum networks such as fiber--coupled single photon sources \cite{ahtee_molecules_2009,hwang_dye_2011,polisseni_stable_2016} or photon sorters \cite{witthaut_photon_2012,ralph_photon_2015}. 

\begin{acknowledgments}
\noindent The scanning electron microscope imaging has been carried out using facilities
at the University Service Centre for Transmission Electron Microscopy (USTEM), Vienna University of Technology, Austria. We gratefully acknowledge financial support by the Austrian Science Fund (FWF Lise Meitner project No. M 2114-N27) and the Federal Ministry of Science, Research and Economy of Austria.
\end{acknowledgments}

\appendix

\section{Lifetime-limited single photons from a fully fiber-integrated single molecule}
\label{app:photons}
To show that we can readily produce fiber-coupled lifetime-limited photons from single molecules in nanocrystals with our set-up, we have measured molecular spectra and corresponding fluorescence correlation measurements at a temperature of 1.7 K. These measurements were performed on a different 
\begin{figure}[h!]
\center
\includegraphics[width=\columnwidth]{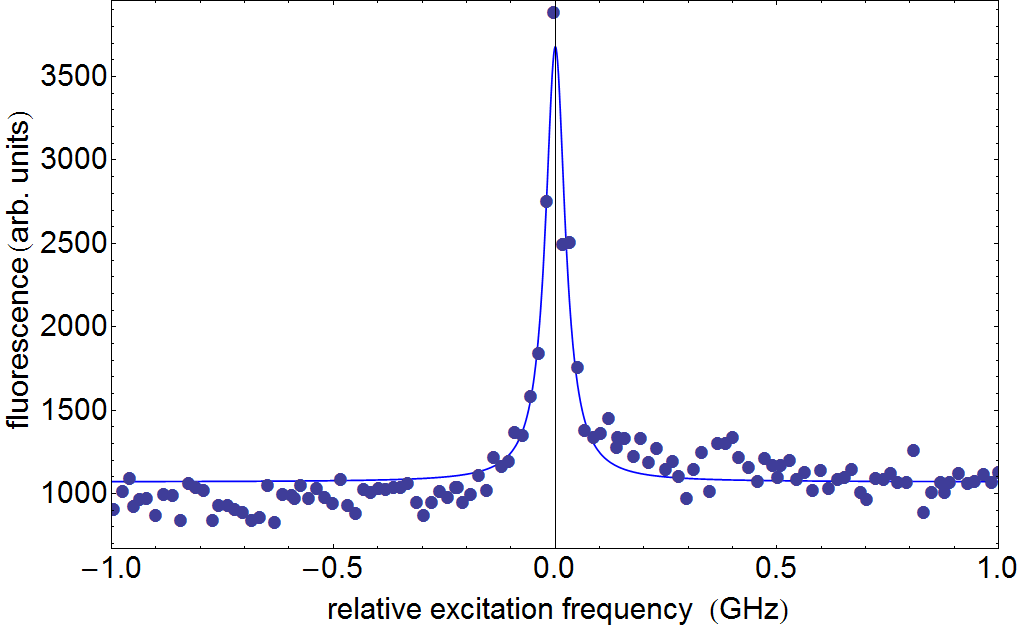}
\caption{Laser induced fluorescence spectrum of a single terrylene molecule coupled to an optical nanofiber at a temperature of 1.7 K}
\label{fig:Mol2KSpectrum}
\end{figure}
\begin{figure}[h!]
\center
\includegraphics[width=\columnwidth]{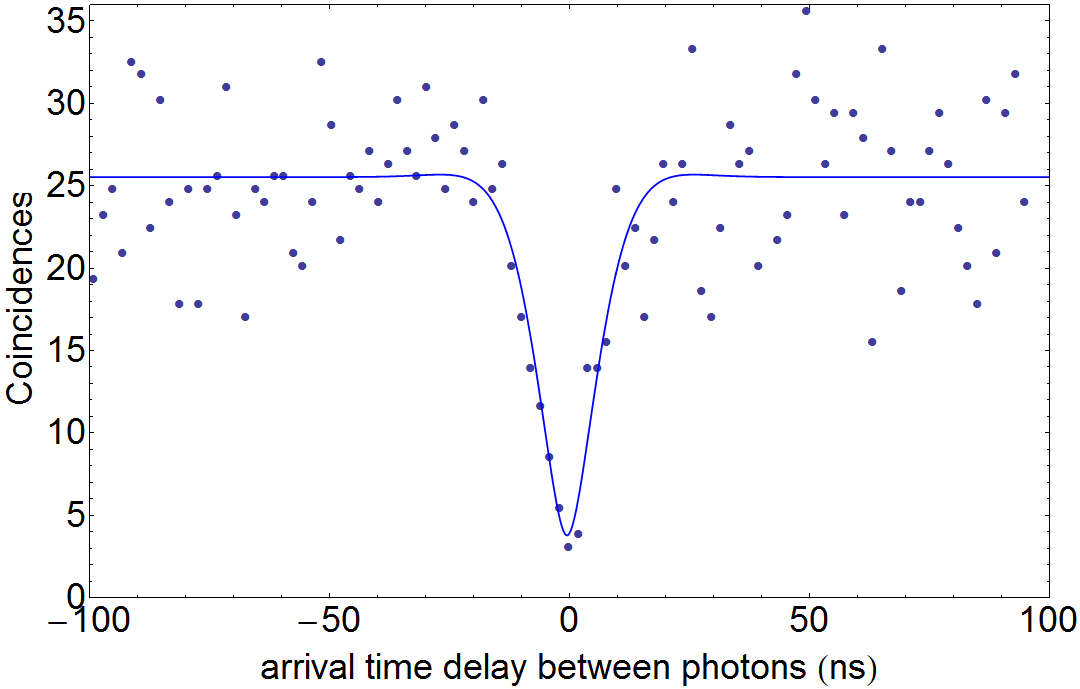}
\caption{Fluorescence intensity correlation measurements of a single terrylene molecule coupled to an optical nanofiber at a temperature of 1.7 K. The coincidences are analysed in bins of 1000 s and the time resolution for the data is 2 ns. }
\label{fig:Mol2Kg2}
\end{figure}
\begin{figure}[htb]
\center
\includegraphics[width=\columnwidth]{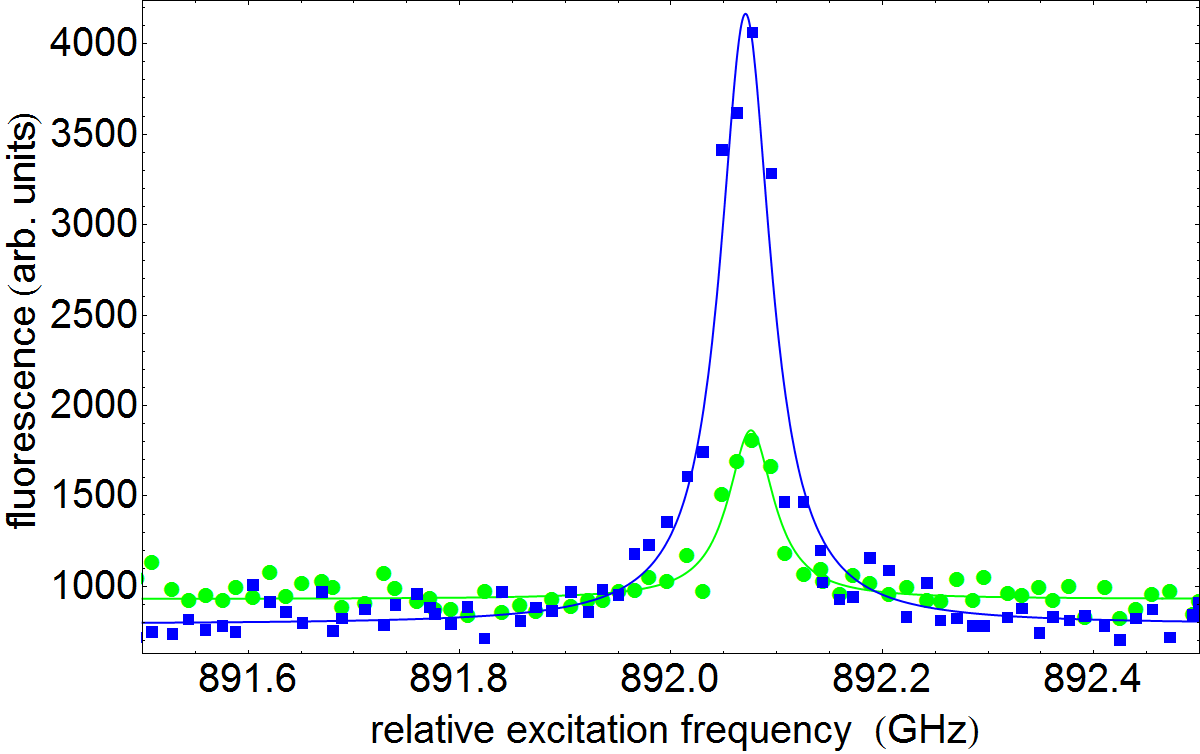}
\caption{Laser induced fluorescence spectra of a single terrylene molecule coupled to an optical nanofiber at a temperature of 1.7 K observed from both sides of the optical fiber. The fit to the spectra reveals a linewidth (FWHM) of $52.5\pm 5.1$ MHz (blue, squares) and $58.6\pm2.7$ MHz (green, circles). The difference in signal height of the two spectra may stem from the possibly asymmetric position of the molecule inside the crystal with respect to the forward and backward propagation direction of laser-induced fluorescence in the optical nanofiber.}
\label{fig:transmissionReflection}
\end{figure}
sample mounted in a helium flow cryostat (Janis) that can cool samples continuously below 2 K. Figure \ref{fig:Mol2KSpectrum} shows a molecular spectrum at 1.7 K that has a linewidth of $54\pm 6$ MHz and Figure \ref{fig:Mol2Kg2} shows the corresponding second order fluorescence correlation measurement yielding a linewidth of $40 \pm 22$ MHz in agreement with that obtained from the spectrum. We can measure such spectra on both ends of the tapered optical fiber as shown in Figure \ref{fig:transmissionReflection}. This underpins the fully fiber-coupled nature of our single molecules.

\section{Excitation efficiency of a molecule on the optical nanofiber surface}
\label{app:excitation}

The efficiency of exciting a molecule via the nanofiber interface is given by $\eta_{\text{abs}} = \sigma/A_{\text{eff}}(x,y) $, where $A_{\text{eff}}(x,y)$ is the effective mode area of the light field at the position of the molecule. Here, we want to estimate the excitation efficiency for a single molecule that is situated on the surface of the optical nanofiber at position $(x,y)=(0,0)$ (see Figure 9 in main text) and has a radially oriented transition dipole moment with respect to the nanofiber. It is excited by a mode that is quasilinearly polarised and for which the transverse polarisation component is aligned with the direction of the molecule's dipole moment. The effective mode area on the surface is then given as $A_{\text{eff,surf}} = P/(I_{\text{surf}}\, (\mathbf{\hat{d}}\cdot \mathbf{\hat{e}_{\text{surf}}})^2)$, where $I_{\text{surf}}$ is the surface intensity at $(0,0)$ and $\mathbf{\hat{d}}\cdot \mathbf{\hat{e}_{\text{surf}}}$ is the overlap between the unit vector of the molecular dipole moment and the polarisation vector at the position of the molecule. On the surface this then yields an effective mode area of 0.4 $\lambda^2$, with an excitation wavelength of 580 nm and for a nanofiber with a diameter of 320 nm. We deduce a value for the absorption cross-section of the molecule from our saturation intensity measurements. The saturation intensity of the molecule closest to the nanofiber has been $I_{\text{S}}<0.5$ Wcm$^{-2}$.  The intensity needed to saturate a molecular transition depends on its molecular dipole moment as \cite{jelezko_read-out_2004}
\begin{equation}
I_{\text{S}} = \epsilon_0 c \hbar^2\frac{k_{21}+k_{23}'+k_{23}''}{\vert\mathbf{d}\vert^2(2+A) T_2}
\label{eq:Ithreelevels}
\end{equation} 
Here, to obtain a correct value for the magnitude of the dipole moment $\mathbf{d}$, the metastable triplet state, which is split into two levels without an external field, has to be taken into account. In this case, $A = k_{23}'/k_{31}'+k_{23}''/k_{31}''$. $k_{ij}$ are the transition rates from level $i$ to level $j$ and the superscripts $'$ and $''$ represent the two non degenerate levels of the triplet level $3$. $1/T_2$ represents the total dephasing rate of the excited state. Using equation \ref{eq:Ithreelevels} and the triplet state parameters for terrylene in p-terphenyl from \cite{hegerfeldt_blinking_2003}, we obtain $\vert\textbf{d}\vert>4.1$ Debye for the transition dipole moment of terrylene in p-terphenyl from the ground state to the first electronically excited state. The molecular dipole moment $\mathbf{d} = -e \mathbf{r}$ yields a minimum oscillator strength for this transition as \cite{book_bransden}:
\begin{equation}
f = \frac{2 m \omega}{3 \hbar} \vert\mathbf{r}\vert^2 = 0.14
\end{equation}
This gives an effective absorption cross-section of $\sigma = 0.14 \times 3 \lambda^2/(2 \pi)$ and hence an excitation probability of a single molecule on the nanofiber surface of 17\%.

\bibliography{quantumBibliography2}

\begin{thebibliography}{84}%
\makeatletter
\providecommand \@ifxundefined [1]{%
 \@ifx{#1\undefined}
}%
\providecommand \@ifnum [1]{%
 \ifnum #1\expandafter \@firstoftwo
 \else \expandafter \@secondoftwo
 \fi
}%
\providecommand \@ifx [1]{%
 \ifx #1\expandafter \@firstoftwo
 \else \expandafter \@secondoftwo
 \fi
}%
\providecommand \natexlab [1]{#1}%
\providecommand \enquote  [1]{``#1''}%
\providecommand \bibnamefont  [1]{#1}%
\providecommand \bibfnamefont [1]{#1}%
\providecommand \citenamefont [1]{#1}%
\providecommand \href@noop [0]{\@secondoftwo}%
\providecommand \href [0]{\begingroup \@sanitize@url \@href}%
\providecommand \@href[1]{\@@startlink{#1}\@@href}%
\providecommand \@@href[1]{\endgroup#1\@@endlink}%
\providecommand \@sanitize@url [0]{\catcode `\\12\catcode `\$12\catcode
  `\&12\catcode `\#12\catcode `\^12\catcode `\_12\catcode `\%12\relax}%
\providecommand \@@startlink[1]{}%
\providecommand \@@endlink[0]{}%
\providecommand \url  [0]{\begingroup\@sanitize@url \@url }%
\providecommand \@url [1]{\endgroup\@href {#1}{\urlprefix }}%
\providecommand \urlprefix  [0]{URL }%
\providecommand \Eprint [0]{\href }%
\providecommand \doibase [0]{http://dx.doi.org/}%
\providecommand \selectlanguage [0]{\@gobble}%
\providecommand \bibinfo  [0]{\@secondoftwo}%
\providecommand \bibfield  [0]{\@secondoftwo}%
\providecommand \translation [1]{[#1]}%
\providecommand \BibitemOpen [0]{}%
\providecommand \bibitemStop [0]{}%
\providecommand \bibitemNoStop [0]{.\EOS\space}%
\providecommand \EOS [0]{\spacefactor3000\relax}%
\providecommand \BibitemShut  [1]{\csname bibitem#1\endcsname}%
\let\auto@bib@innerbib\@empty
\bibitem [{\citenamefont {Siyushev}\ \emph {et~al.}(2014)\citenamefont
  {Siyushev}, \citenamefont {Stein}, \citenamefont {Wrachtrup},\ and\
  \citenamefont {Gerhardt}}]{siyushev_molecular_2014}%
  \BibitemOpen
  \bibfield  {author} {\bibinfo {author} {\bibfnamefont {P.}~\bibnamefont
  {Siyushev}}, \bibinfo {author} {\bibfnamefont {G.}~\bibnamefont {Stein}},
  \bibinfo {author} {\bibfnamefont {J.}~\bibnamefont {Wrachtrup}}, \ and\
  \bibinfo {author} {\bibfnamefont {I.}~\bibnamefont {Gerhardt}},\ }\bibfield
  {title} {{\selectlanguage {english}\enquote {\bibinfo {title} {Molecular
  photons interfaced with alkali atoms},}\ }}\href {\doibase
  10.1038/nature13191} {\bibfield  {journal} {\bibinfo  {journal} {Nature}\
  }\textbf {\bibinfo {volume} {509}},\ \bibinfo {pages} {66--70} (\bibinfo
  {year} {2014})}\BibitemShut {NoStop}%
\bibitem [{\citenamefont {Basch\'{e}}\ \emph {et~al.}(1995)\citenamefont
  {Basch\'{e}}, \citenamefont {Kummer},\ and\ \citenamefont
  {Br\"{a}uchle}}]{basche_direct_1995}%
  \BibitemOpen
  \bibfield  {author} {\bibinfo {author} {\bibfnamefont {Th}~\bibnamefont
  {Basch\'{e}}}, \bibinfo {author} {\bibfnamefont {S.}~\bibnamefont {Kummer}},
  \ and\ \bibinfo {author} {\bibfnamefont {C.}~\bibnamefont {Br\"{a}uchle}},\
  }\bibfield  {title} {{\selectlanguage {english}\enquote {\bibinfo {title}
  {Direct spectroscopic observation of quantum jumps of a single molecule},}\
  }}\href {\doibase 10.1038/373132a0} {\bibfield  {journal} {\bibinfo
  {journal} {Nature}\ }\textbf {\bibinfo {volume} {373}},\ \bibinfo {pages}
  {132--134} (\bibinfo {year} {1995})}\BibitemShut {NoStop}%
\bibitem [{\citenamefont {Basch\'{e}}\ \emph {et~al.}(1992)\citenamefont
  {Basch\'{e}}, \citenamefont {Moerner}, \citenamefont {Orrit},\ and\
  \citenamefont {Talon}}]{basche_photon_1992}%
  \BibitemOpen
  \bibfield  {author} {\bibinfo {author} {\bibfnamefont {Th.}\ \bibnamefont
  {Basch\'{e}}}, \bibinfo {author} {\bibfnamefont {W.~E.}\ \bibnamefont
  {Moerner}}, \bibinfo {author} {\bibfnamefont {M.}~\bibnamefont {Orrit}}, \
  and\ \bibinfo {author} {\bibfnamefont {H.}~\bibnamefont {Talon}},\ }\bibfield
   {title} {\enquote {\bibinfo {title} {Photon antibunching in the fluorescence
  of a single dye molecule trapped in a solid},}\ }\href {\doibase
  10.1103/PhysRevLett.69.1516} {\bibfield  {journal} {\bibinfo  {journal}
  {Phys. Rev. Lett.}\ }\textbf {\bibinfo {volume} {69}},\ \bibinfo {pages}
  {1516--1519} (\bibinfo {year} {1992})}\BibitemShut {NoStop}%
\bibitem [{\citenamefont {Celebrano}\ \emph {et~al.}(2011)\citenamefont
  {Celebrano}, \citenamefont {Kukura}, \citenamefont {Renn},\ and\
  \citenamefont {Sandoghdar}}]{celebrano_single-molecule_2011}%
  \BibitemOpen
  \bibfield  {author} {\bibinfo {author} {\bibfnamefont {M.}~\bibnamefont
  {Celebrano}}, \bibinfo {author} {\bibfnamefont {P.}~\bibnamefont {Kukura}},
  \bibinfo {author} {\bibfnamefont {A.}~\bibnamefont {Renn}}, \ and\ \bibinfo
  {author} {\bibfnamefont {V.}~\bibnamefont {Sandoghdar}},\ }\bibfield  {title}
  {\enquote {\bibinfo {title} {Single-molecule imaging by optical
  absorption},}\ }\href {\doibase 10.1038/nphoton.2010.290} {\bibfield
  {journal} {\bibinfo  {journal} {Nat. Photon.}\ }\textbf {\bibinfo {volume}
  {5}},\ \bibinfo {pages} {95} (\bibinfo {year} {2011})}\BibitemShut {NoStop}%
\bibitem [{\citenamefont {Gerhardt}\ \emph {et~al.}(2007)\citenamefont
  {Gerhardt}, \citenamefont {Wrigge}, \citenamefont {Bushev}, \citenamefont
  {Zumofen}, \citenamefont {Agio}, \citenamefont {Pfab},\ and\ \citenamefont
  {Sandoghdar}}]{gerhardt_strong_2007}%
  \BibitemOpen
  \bibfield  {author} {\bibinfo {author} {\bibfnamefont {I.}~\bibnamefont
  {Gerhardt}}, \bibinfo {author} {\bibfnamefont {G.}~\bibnamefont {Wrigge}},
  \bibinfo {author} {\bibfnamefont {P.}~\bibnamefont {Bushev}}, \bibinfo
  {author} {\bibfnamefont {G.}~\bibnamefont {Zumofen}}, \bibinfo {author}
  {\bibfnamefont {M.}~\bibnamefont {Agio}}, \bibinfo {author} {\bibfnamefont
  {R.}~\bibnamefont {Pfab}}, \ and\ \bibinfo {author} {\bibfnamefont
  {V.}~\bibnamefont {Sandoghdar}},\ }\bibfield  {title} {\enquote {\bibinfo
  {title} {Strong {Extinction} of a {Laser} {Beam} by a {Single} {Molecule}},}\
  }\href {\doibase 10.1103/PhysRevLett.98.033601} {\bibfield  {journal}
  {\bibinfo  {journal} {Phys. Rev. Lett.}\ }\textbf {\bibinfo {volume} {98}},\
  \bibinfo {pages} {033601} (\bibinfo {year} {2007})}\BibitemShut {NoStop}%
\bibitem [{\citenamefont {Gerhardt}\ \emph {et~al.}(2010)\citenamefont
  {Gerhardt}, \citenamefont {Wrigge}, \citenamefont {Hwang}, \citenamefont
  {Zumofen},\ and\ \citenamefont {Sandoghdar}}]{gerhardt_coherent_2010}%
  \BibitemOpen
  \bibfield  {author} {\bibinfo {author} {\bibfnamefont {I.}~\bibnamefont
  {Gerhardt}}, \bibinfo {author} {\bibfnamefont {G.}~\bibnamefont {Wrigge}},
  \bibinfo {author} {\bibfnamefont {J.}~\bibnamefont {Hwang}}, \bibinfo
  {author} {\bibfnamefont {G.}~\bibnamefont {Zumofen}}, \ and\ \bibinfo
  {author} {\bibfnamefont {V.}~\bibnamefont {Sandoghdar}},\ }\bibfield  {title}
  {\enquote {\bibinfo {title} {Coherent nonlinear single-molecule
  microscopy},}\ }\href {\doibase 10.1103/PhysRevA.82.063823} {\bibfield
  {journal} {\bibinfo  {journal} {Phys. Rev. A}\ }\textbf {\bibinfo {volume}
  {82}},\ \bibinfo {pages} {063823} (\bibinfo {year} {2010})}\BibitemShut
  {NoStop}%
\bibitem [{\citenamefont {Kozankiewicz}\ and\ \citenamefont
  {Orrit}(2014)}]{kozankiewicz_single-molecule_2014}%
  \BibitemOpen
  \bibfield  {author} {\bibinfo {author} {\bibfnamefont {B.}~\bibnamefont
  {Kozankiewicz}}\ and\ \bibinfo {author} {\bibfnamefont {M.}~\bibnamefont
  {Orrit}},\ }\bibfield  {title} {{\selectlanguage {english}\enquote {\bibinfo
  {title} {Single-molecule photophysics, from cryogenic to ambient
  conditions},}\ }}\href {\doibase 10.1039/C3CS60165J} {\bibfield  {journal}
  {\bibinfo  {journal} {Chem. Soc. Rev.}\ }\textbf {\bibinfo {volume} {43}},\
  \bibinfo {pages} {1029--1043} (\bibinfo {year} {2014})}\BibitemShut {NoStop}%
\bibitem [{\citenamefont {Jelezko}\ and\ \citenamefont
  {Wrachtrup}(2004)}]{jelezko_read-out_2004}%
  \BibitemOpen
  \bibfield  {author} {\bibinfo {author} {\bibfnamefont {F.}~\bibnamefont
  {Jelezko}}\ and\ \bibinfo {author} {\bibfnamefont {J.}~\bibnamefont
  {Wrachtrup}},\ }\bibfield  {title} {{\selectlanguage {english}\enquote
  {\bibinfo {title} {Read-out of single spins by optical spectroscopy},}\
  }}\href {\doibase 10.1088/0953-8984/16/30/R03} {\bibfield  {journal}
  {\bibinfo  {journal} {J. Phys.: Condens. Matter}\ }\textbf {\bibinfo {volume}
  {16}},\ \bibinfo {pages} {R1089} (\bibinfo {year} {2004})}\BibitemShut
  {NoStop}%
\bibitem [{\citenamefont {Maurer}\ \emph {et~al.}(2012)\citenamefont {Maurer},
  \citenamefont {Kucsko}, \citenamefont {Latta}, \citenamefont {Jiang},
  \citenamefont {Yao}, \citenamefont {Bennett}, \citenamefont {Pastawski},
  \citenamefont {Hunger}, \citenamefont {Chisholm}, \citenamefont {Markham},
  \citenamefont {Twitchen}, \citenamefont {Cirac},\ and\ \citenamefont
  {Lukin}}]{maurer_room-temperature_2012}%
  \BibitemOpen
  \bibfield  {author} {\bibinfo {author} {\bibfnamefont {P.~C.}\ \bibnamefont
  {Maurer}}, \bibinfo {author} {\bibfnamefont {G.}~\bibnamefont {Kucsko}},
  \bibinfo {author} {\bibfnamefont {C.}~\bibnamefont {Latta}}, \bibinfo
  {author} {\bibfnamefont {L.}~\bibnamefont {Jiang}}, \bibinfo {author}
  {\bibfnamefont {N.~Y.}\ \bibnamefont {Yao}}, \bibinfo {author} {\bibfnamefont
  {S.~D.}\ \bibnamefont {Bennett}}, \bibinfo {author} {\bibfnamefont
  {F.}~\bibnamefont {Pastawski}}, \bibinfo {author} {\bibfnamefont
  {D.}~\bibnamefont {Hunger}}, \bibinfo {author} {\bibfnamefont
  {N.}~\bibnamefont {Chisholm}}, \bibinfo {author} {\bibfnamefont
  {M.}~\bibnamefont {Markham}}, \bibinfo {author} {\bibfnamefont {D.~J.}\
  \bibnamefont {Twitchen}}, \bibinfo {author} {\bibfnamefont {J.~I.}\
  \bibnamefont {Cirac}}, \ and\ \bibinfo {author} {\bibfnamefont {M.~D.}\
  \bibnamefont {Lukin}},\ }\bibfield  {title} {{\selectlanguage
  {english}\enquote {\bibinfo {title} {Room-{Temperature} {Quantum} {Bit}
  {Memory} {Exceeding} {One} {Second}},}\ }}\href {\doibase
  10.1126/science.1220513} {\bibfield  {journal} {\bibinfo  {journal}
  {Science}\ }\textbf {\bibinfo {volume} {336}},\ \bibinfo {pages} {1283--1286}
  (\bibinfo {year} {2012})}\BibitemShut {NoStop}%
\bibitem [{\citenamefont {Neumann}\ \emph {et~al.}(2008)\citenamefont
  {Neumann}, \citenamefont {Mizuochi}, \citenamefont {Rempp}, \citenamefont
  {Hemmer}, \citenamefont {Watanabe}, \citenamefont {Yamasaki}, \citenamefont
  {Jacques}, \citenamefont {Gaebel}, \citenamefont {Jelezko},\ and\
  \citenamefont {Wrachtrup}}]{neumann_multipartite_2008}%
  \BibitemOpen
  \bibfield  {author} {\bibinfo {author} {\bibfnamefont {P.}~\bibnamefont
  {Neumann}}, \bibinfo {author} {\bibfnamefont {N.}~\bibnamefont {Mizuochi}},
  \bibinfo {author} {\bibfnamefont {F.}~\bibnamefont {Rempp}}, \bibinfo
  {author} {\bibfnamefont {P.}~\bibnamefont {Hemmer}}, \bibinfo {author}
  {\bibfnamefont {H.}~\bibnamefont {Watanabe}}, \bibinfo {author}
  {\bibfnamefont {S.}~\bibnamefont {Yamasaki}}, \bibinfo {author}
  {\bibfnamefont {V.}~\bibnamefont {Jacques}}, \bibinfo {author} {\bibfnamefont
  {T.}~\bibnamefont {Gaebel}}, \bibinfo {author} {\bibfnamefont
  {F.}~\bibnamefont {Jelezko}}, \ and\ \bibinfo {author} {\bibfnamefont
  {J.}~\bibnamefont {Wrachtrup}},\ }\bibfield  {title} {{\selectlanguage
  {english}\enquote {\bibinfo {title} {Multipartite {Entanglement} {Among}
  {Single} {Spins} in {Diamond}},}\ }}\href {\doibase 10.1126/science.1157233}
  {\bibfield  {journal} {\bibinfo  {journal} {Science}\ }\textbf {\bibinfo
  {volume} {320}},\ \bibinfo {pages} {1326--1329} (\bibinfo {year}
  {2008})}\BibitemShut {NoStop}%
\bibitem [{\citenamefont {Pingault}\ \emph {et~al.}(2014)\citenamefont
  {Pingault}, \citenamefont {Becker}, \citenamefont {Schulte}, \citenamefont
  {Arend}, \citenamefont {Hepp}, \citenamefont {Godde}, \citenamefont
  {Tartakovskii}, \citenamefont {Markham}, \citenamefont {Becher},\ and\
  \citenamefont {Atat\"{u}re}}]{pingault_all-optical_2014}%
  \BibitemOpen
  \bibfield  {author} {\bibinfo {author} {\bibfnamefont {B.}~\bibnamefont
  {Pingault}}, \bibinfo {author} {\bibfnamefont {J.~N.}\ \bibnamefont
  {Becker}}, \bibinfo {author} {\bibfnamefont {C.~H.~H.}\ \bibnamefont
  {Schulte}}, \bibinfo {author} {\bibfnamefont {C.}~\bibnamefont {Arend}},
  \bibinfo {author} {\bibfnamefont {C.}~\bibnamefont {Hepp}}, \bibinfo {author}
  {\bibfnamefont {T.}~\bibnamefont {Godde}}, \bibinfo {author} {\bibfnamefont
  {A.~I.}\ \bibnamefont {Tartakovskii}}, \bibinfo {author} {\bibfnamefont
  {M.}~\bibnamefont {Markham}}, \bibinfo {author} {\bibfnamefont
  {C.}~\bibnamefont {Becher}}, \ and\ \bibinfo {author} {\bibfnamefont
  {M.}~\bibnamefont {Atat\"{u}re}},\ }\bibfield  {title} {\enquote {\bibinfo
  {title} {All-{Optical} {Formation} of {Coherent} {Dark} {States} of
  {Silicon}-{Vacancy} {Spins} in {Diamond}},}\ }\href {\doibase
  10.1103/PhysRevLett.113.263601} {\bibfield  {journal} {\bibinfo  {journal}
  {Phys. Rev. Lett.}\ }\textbf {\bibinfo {volume} {113}},\ \bibinfo {pages}
  {263601} (\bibinfo {year} {2014})}\BibitemShut {NoStop}%
\bibitem [{\citenamefont {Hansom}\ \emph
  {et~al.}(2014{\natexlab{a}})\citenamefont {Hansom}, \citenamefont {Schulte},
  \citenamefont {Le~Gall}, \citenamefont {Matthiesen}, \citenamefont {Clarke},
  \citenamefont {Hugues}, \citenamefont {Taylor},\ and\ \citenamefont
  {Atat\"{u}re}}]{hansom_environment-assisted_2014}%
  \BibitemOpen
  \bibfield  {author} {\bibinfo {author} {\bibfnamefont {J.}~\bibnamefont
  {Hansom}}, \bibinfo {author} {\bibfnamefont {C.~H.~H.}\ \bibnamefont
  {Schulte}}, \bibinfo {author} {\bibfnamefont {C.}~\bibnamefont {Le~Gall}},
  \bibinfo {author} {\bibfnamefont {C.}~\bibnamefont {Matthiesen}}, \bibinfo
  {author} {\bibfnamefont {E.}~\bibnamefont {Clarke}}, \bibinfo {author}
  {\bibfnamefont {M.}~\bibnamefont {Hugues}}, \bibinfo {author} {\bibfnamefont
  {J.~M.}\ \bibnamefont {Taylor}}, \ and\ \bibinfo {author} {\bibfnamefont
  {M.}~\bibnamefont {Atat\"{u}re}},\ }\bibfield  {title} {{\selectlanguage
  {english}\enquote {\bibinfo {title} {Environment-assisted quantum control of
  a solid-state spin via coherent dark states},}\ }}\href {\doibase
  10.1038/nphys3077} {\bibfield  {journal} {\bibinfo  {journal} {Nat. Phys.}\
  }\textbf {\bibinfo {volume} {10}},\ \bibinfo {pages} {725--730} (\bibinfo
  {year} {2014}{\natexlab{a}})}\BibitemShut {NoStop}%
\bibitem [{\citenamefont {Delteil}\ \emph {et~al.}(2016)\citenamefont
  {Delteil}, \citenamefont {Sun}, \citenamefont {Gao}, \citenamefont {Togan},
  \citenamefont {Faelt},\ and\ \citenamefont
  {Imamo\u{g}lu}}]{delteil_generation_2016}%
  \BibitemOpen
  \bibfield  {author} {\bibinfo {author} {\bibfnamefont {A.}~\bibnamefont
  {Delteil}}, \bibinfo {author} {\bibfnamefont {Z.}~\bibnamefont {Sun}},
  \bibinfo {author} {\bibfnamefont {W.}~\bibnamefont {Gao}}, \bibinfo {author}
  {\bibfnamefont {E.}~\bibnamefont {Togan}}, \bibinfo {author} {\bibfnamefont
  {S.}~\bibnamefont {Faelt}}, \ and\ \bibinfo {author} {\bibfnamefont
  {A.}~\bibnamefont {Imamo\u{g}lu}},\ }\bibfield  {title} {{\selectlanguage
  {english}\enquote {\bibinfo {title} {Generation of heralded entanglement
  between distant hole spins},}\ }}\href {\doibase 10.1038/nphys3605}
  {\bibfield  {journal} {\bibinfo  {journal} {Nat. Phys.}\ }\textbf {\bibinfo
  {volume} {12}},\ \bibinfo {pages} {218--223} (\bibinfo {year}
  {2016})}\BibitemShut {NoStop}%
\bibitem [{\citenamefont {Yalla}\ \emph {et~al.}(2014)\citenamefont {Yalla},
  \citenamefont {Sadgrove}, \citenamefont {Nayak},\ and\ \citenamefont
  {Hakuta}}]{Yalla_cavity_2014}%
  \BibitemOpen
  \bibfield  {author} {\bibinfo {author} {\bibfnamefont {R.}~\bibnamefont
  {Yalla}}, \bibinfo {author} {\bibfnamefont {M.}~\bibnamefont {Sadgrove}},
  \bibinfo {author} {\bibfnamefont {K.~P.}\ \bibnamefont {Nayak}}, \ and\
  \bibinfo {author} {\bibfnamefont {K.}~\bibnamefont {Hakuta}},\ }\bibfield
  {title} {\enquote {\bibinfo {title} {Cavity quantum electrodynamics on a
  nanofiber using a composite photonic crystal cavity},}\ }\href@noop {}
  {\bibfield  {journal} {\bibinfo  {journal} {Phys. Rev. Lett.}\ }\textbf
  {\bibinfo {volume} {113}},\ \bibinfo {pages} {143601} (\bibinfo {year}
  {2014})}\BibitemShut {NoStop}%
\bibitem [{\citenamefont {Arcari}\ \emph {et~al.}(2014)\citenamefont {Arcari},
  \citenamefont {S\"{o}llner}, \citenamefont {Javadi}, \citenamefont
  {Lindskov~Hansen}, \citenamefont {Mahmoodian}, \citenamefont {Liu},
  \citenamefont {Thyrrestrup}, \citenamefont {Lee}, \citenamefont {Song},
  \citenamefont {Stobbe},\ and\ \citenamefont {Lodahl}}]{Arcari_photonic_2014}%
  \BibitemOpen
  \bibfield  {author} {\bibinfo {author} {\bibfnamefont {M.}~\bibnamefont
  {Arcari}}, \bibinfo {author} {\bibfnamefont {I.}~\bibnamefont {S\"{o}llner}},
  \bibinfo {author} {\bibfnamefont {A.}~\bibnamefont {Javadi}}, \bibinfo
  {author} {\bibfnamefont {S.}~\bibnamefont {Lindskov~Hansen}}, \bibinfo
  {author} {\bibfnamefont {S.}~\bibnamefont {Mahmoodian}}, \bibinfo {author}
  {\bibfnamefont {J.}~\bibnamefont {Liu}}, \bibinfo {author} {\bibfnamefont
  {H.}~\bibnamefont {Thyrrestrup}}, \bibinfo {author} {\bibfnamefont {E.~H.}\
  \bibnamefont {Lee}}, \bibinfo {author} {\bibfnamefont {J.~D.}\ \bibnamefont
  {Song}}, \bibinfo {author} {\bibfnamefont {S.}~\bibnamefont {Stobbe}}, \ and\
  \bibinfo {author} {\bibfnamefont {P.}~\bibnamefont {Lodahl}},\ }\bibfield
  {title} {\enquote {\bibinfo {title} {Near-unity coupling efficiency of a
  quantum emitter to a photonic crystal waveguide},}\ }\href@noop {} {\bibfield
   {journal} {\bibinfo  {journal} {Phys. Rev. Lett.}\ }\textbf {\bibinfo
  {volume} {113}},\ \bibinfo {pages} {093603} (\bibinfo {year}
  {2014})}\BibitemShut {NoStop}%
\bibitem [{\citenamefont {Javadi}\ \emph {et~al.}(2015)\citenamefont {Javadi},
  \citenamefont {S\"{o}llner}, \citenamefont {Arcari}, \citenamefont
  {Lindskov~Hansen}, \citenamefont {Midolo}, \citenamefont {Mahmoodian},
  \citenamefont {Kir\u{s}ansk\.{e}}, \citenamefont {Pregnolato}, \citenamefont
  {Lee}, \citenamefont {Song}, \citenamefont {Stobbe},\ and\ \citenamefont
  {Lodahl}}]{Javadi_waveguide_2015}%
  \BibitemOpen
  \bibfield  {author} {\bibinfo {author} {\bibfnamefont {A.}~\bibnamefont
  {Javadi}}, \bibinfo {author} {\bibfnamefont {I.}~\bibnamefont {S\"{o}llner}},
  \bibinfo {author} {\bibfnamefont {M.}~\bibnamefont {Arcari}}, \bibinfo
  {author} {\bibfnamefont {S.}~\bibnamefont {Lindskov~Hansen}}, \bibinfo
  {author} {\bibfnamefont {L.}~\bibnamefont {Midolo}}, \bibinfo {author}
  {\bibfnamefont {S.}~\bibnamefont {Mahmoodian}}, \bibinfo {author}
  {\bibfnamefont {G.}~\bibnamefont {Kir\u{s}ansk\.{e}}}, \bibinfo {author}
  {\bibfnamefont {T.}~\bibnamefont {Pregnolato}}, \bibinfo {author}
  {\bibfnamefont {E.~H.}\ \bibnamefont {Lee}}, \bibinfo {author} {\bibfnamefont
  {J.~D.}\ \bibnamefont {Song}}, \bibinfo {author} {\bibfnamefont
  {S.}~\bibnamefont {Stobbe}}, \ and\ \bibinfo {author} {\bibfnamefont
  {P.}~\bibnamefont {Lodahl}},\ }\bibfield  {title} {\enquote {\bibinfo {title}
  {Single-photon non-linear optics with a quantum dot in a waveguide},}\
  }\href@noop {} {\bibfield  {journal} {\bibinfo  {journal} {Nat. Commun.}\
  }\textbf {\bibinfo {volume} {6}},\ \bibinfo {pages} {8655} (\bibinfo {year}
  {2015})}\BibitemShut {NoStop}%
\bibitem [{\citenamefont {Nemoto}\ \emph {et~al.}(2014)\citenamefont {Nemoto},
  \citenamefont {Trupke}, \citenamefont {Devitt}, \citenamefont {Stephens},
  \citenamefont {Scharfenberger}, \citenamefont {Buczak}, \citenamefont
  {N\"{o}bauer}, \citenamefont {Everitt}, \citenamefont {Schmiedmayer},\ and\
  \citenamefont {Munro}}]{nemoto_photonic_2014}%
  \BibitemOpen
  \bibfield  {author} {\bibinfo {author} {\bibfnamefont {K.}~\bibnamefont
  {Nemoto}}, \bibinfo {author} {\bibfnamefont {M.}~\bibnamefont {Trupke}},
  \bibinfo {author} {\bibfnamefont {S.~J.}\ \bibnamefont {Devitt}}, \bibinfo
  {author} {\bibfnamefont {A.~M.}\ \bibnamefont {Stephens}}, \bibinfo {author}
  {\bibfnamefont {B.}~\bibnamefont {Scharfenberger}}, \bibinfo {author}
  {\bibfnamefont {K.}~\bibnamefont {Buczak}}, \bibinfo {author} {\bibfnamefont
  {T.}~\bibnamefont {N\"{o}bauer}}, \bibinfo {author} {\bibfnamefont {M.~S.}\
  \bibnamefont {Everitt}}, \bibinfo {author} {\bibfnamefont {J.}~\bibnamefont
  {Schmiedmayer}}, \ and\ \bibinfo {author} {\bibfnamefont {W.~J.}\
  \bibnamefont {Munro}},\ }\bibfield  {title} {\enquote {\bibinfo {title}
  {Photonic {Architecture} for {Scalable} {Quantum} {Information} {Processing}
  in {Diamond}},}\ }\href {\doibase 10.1103/PhysRevX.4.031022} {\bibfield
  {journal} {\bibinfo  {journal} {Phys. Rev. X}\ }\textbf {\bibinfo {volume}
  {4}},\ \bibinfo {pages} {031022} (\bibinfo {year} {2014})}\BibitemShut
  {NoStop}%
\bibitem [{\citenamefont {Kimble}(2008)}]{kimble_quantum_2008}%
  \BibitemOpen
  \bibfield  {author} {\bibinfo {author} {\bibfnamefont {H.~J.}\ \bibnamefont
  {Kimble}},\ }\bibfield  {title} {\enquote {\bibinfo {title} {The quantum
  internet},}\ }\href {\doibase 10.1038/nature07127} {\bibfield  {journal}
  {\bibinfo  {journal} {Nature}\ }\textbf {\bibinfo {volume} {453}},\ \bibinfo
  {pages} {1023--1030} (\bibinfo {year} {2008})}\BibitemShut {NoStop}%
\bibitem [{\citenamefont {Doherty}\ \emph {et~al.}(2014)\citenamefont
  {Doherty}, \citenamefont {Struzhkin}, \citenamefont {Simpson}, \citenamefont
  {McGuinness}, \citenamefont {Meng}, \citenamefont {Stacey}, \citenamefont
  {Karle}, \citenamefont {Hemley}, \citenamefont {Manson}, \citenamefont
  {Hollenberg},\ and\ \citenamefont {Prawer}}]{doherty_electronic_2014}%
  \BibitemOpen
  \bibfield  {author} {\bibinfo {author} {\bibfnamefont {M.~W.}\ \bibnamefont
  {Doherty}}, \bibinfo {author} {\bibfnamefont {V.~V.}\ \bibnamefont
  {Struzhkin}}, \bibinfo {author} {\bibfnamefont {D.~A.}\ \bibnamefont
  {Simpson}}, \bibinfo {author} {\bibfnamefont {L.~P.}\ \bibnamefont
  {McGuinness}}, \bibinfo {author} {\bibfnamefont {Y.}~\bibnamefont {Meng}},
  \bibinfo {author} {\bibfnamefont {A.}~\bibnamefont {Stacey}}, \bibinfo
  {author} {\bibfnamefont {T.~J.}\ \bibnamefont {Karle}}, \bibinfo {author}
  {\bibfnamefont {R.~J.}\ \bibnamefont {Hemley}}, \bibinfo {author}
  {\bibfnamefont {N.~B.}\ \bibnamefont {Manson}}, \bibinfo {author}
  {\bibfnamefont {Lloyd C.~L.}\ \bibnamefont {Hollenberg}}, \ and\ \bibinfo
  {author} {\bibfnamefont {S.}~\bibnamefont {Prawer}},\ }\bibfield  {title}
  {\enquote {\bibinfo {title} {Electronic {Properties} and {Metrology}
  {Applications} of the {Diamond} {NV}$^-$ {Center} under {Pressure}},}\ }\href
  {\doibase 10.1103/PhysRevLett.112.047601} {\bibfield  {journal} {\bibinfo
  {journal} {Phys. Rev. Lett.}\ }\textbf {\bibinfo {volume} {112}},\ \bibinfo
  {pages} {047601} (\bibinfo {year} {2014})}\BibitemShut {NoStop}%
\bibitem [{\citenamefont {Goldstein}\ \emph {et~al.}(2011)\citenamefont
  {Goldstein}, \citenamefont {Cappellaro}, \citenamefont {Maze}, \citenamefont
  {Hodges}, \citenamefont {Jiang}, \citenamefont {S{\o}rensen},\ and\
  \citenamefont {Lukin}}]{goldstein_environment-assisted_2011}%
  \BibitemOpen
  \bibfield  {author} {\bibinfo {author} {\bibfnamefont {G.}~\bibnamefont
  {Goldstein}}, \bibinfo {author} {\bibfnamefont {P.}~\bibnamefont
  {Cappellaro}}, \bibinfo {author} {\bibfnamefont {J.~R.}\ \bibnamefont
  {Maze}}, \bibinfo {author} {\bibfnamefont {J.~S.}\ \bibnamefont {Hodges}},
  \bibinfo {author} {\bibfnamefont {L.}~\bibnamefont {Jiang}}, \bibinfo
  {author} {\bibfnamefont {A.~S.}\ \bibnamefont {S{\o}rensen}}, \ and\ \bibinfo
  {author} {\bibfnamefont {M.~D.}\ \bibnamefont {Lukin}},\ }\bibfield  {title}
  {\enquote {\bibinfo {title} {Environment-{Assisted} {Precision}
  {Measurement}},}\ }\href {\doibase 10.1103/PhysRevLett.106.140502} {\bibfield
   {journal} {\bibinfo  {journal} {Phys. Rev. Lett.}\ }\textbf {\bibinfo
  {volume} {106}},\ \bibinfo {pages} {140502} (\bibinfo {year}
  {2011})}\BibitemShut {NoStop}%
\bibitem [{\citenamefont {Acosta}\ \emph {et~al.}(2013)\citenamefont {Acosta},
  \citenamefont {Jensen}, \citenamefont {Santori}, \citenamefont {Budker},\
  and\ \citenamefont {Beausoleil}}]{acosta_electromagnetically_2013}%
  \BibitemOpen
  \bibfield  {author} {\bibinfo {author} {\bibfnamefont {V.~M.}\ \bibnamefont
  {Acosta}}, \bibinfo {author} {\bibfnamefont {K.}~\bibnamefont {Jensen}},
  \bibinfo {author} {\bibfnamefont {C.}~\bibnamefont {Santori}}, \bibinfo
  {author} {\bibfnamefont {D.}~\bibnamefont {Budker}}, \ and\ \bibinfo {author}
  {\bibfnamefont {R.~G.}\ \bibnamefont {Beausoleil}},\ }\bibfield  {title}
  {\enquote {\bibinfo {title} {Electromagnetically {Induced} {Transparency} in
  a {Diamond} {Spin} {Ensemble} {Enables} {All}-{Optical} {Electromagnetic}
  {Field} {Sensing}},}\ }\href {\doibase 10.1103/PhysRevLett.110.213605}
  {\bibfield  {journal} {\bibinfo  {journal} {Phys. Rev. Lett.}\ }\textbf
  {\bibinfo {volume} {110}},\ \bibinfo {pages} {213605} (\bibinfo {year}
  {2013})}\BibitemShut {NoStop}%
\bibitem [{\citenamefont {Faez}\ \emph
  {et~al.}(2014{\natexlab{a}})\citenamefont {Faez}, \citenamefont {van~der
  Molen},\ and\ \citenamefont {Orrit}}]{faez_optical_2014}%
  \BibitemOpen
  \bibfield  {author} {\bibinfo {author} {\bibfnamefont {S.}~\bibnamefont
  {Faez}}, \bibinfo {author} {\bibfnamefont {S.~J.}\ \bibnamefont {van~der
  Molen}}, \ and\ \bibinfo {author} {\bibfnamefont {M.}~\bibnamefont {Orrit}},\
  }\bibfield  {title} {\enquote {\bibinfo {title} {Optical tracing of multiple
  charges in single-electron devices},}\ }\href {\doibase
  10.1103/PhysRevB.90.205405} {\bibfield  {journal} {\bibinfo  {journal} {Phys.
  Rev. B}\ }\textbf {\bibinfo {volume} {90}},\ \bibinfo {pages} {205405}
  (\bibinfo {year} {2014}{\natexlab{a}})}\BibitemShut {NoStop}%
\bibitem [{\citenamefont {Mazzamuto}\ \emph {et~al.}(2014)\citenamefont
  {Mazzamuto}, \citenamefont {Tabani}, \citenamefont {Pazzagli}, \citenamefont
  {Rizvi}, \citenamefont {Reserbat-Plantey}, \citenamefont {Sch\"{a}dler},
  \citenamefont {Navickaite}, \citenamefont {Gaudreau}, \citenamefont
  {Cataliotti}, \citenamefont {Koppens},\ and\ \citenamefont
  {Toninelli}}]{mazzamuto_single-molecule_2014}%
  \BibitemOpen
  \bibfield  {author} {\bibinfo {author} {\bibfnamefont {G.}~\bibnamefont
  {Mazzamuto}}, \bibinfo {author} {\bibfnamefont {A.}~\bibnamefont {Tabani}},
  \bibinfo {author} {\bibfnamefont {S.}~\bibnamefont {Pazzagli}}, \bibinfo
  {author} {\bibfnamefont {S.}~\bibnamefont {Rizvi}}, \bibinfo {author}
  {\bibfnamefont {A.}~\bibnamefont {Reserbat-Plantey}}, \bibinfo {author}
  {\bibfnamefont {K.}~\bibnamefont {Sch\"{a}dler}}, \bibinfo {author}
  {\bibfnamefont {G.}~\bibnamefont {Navickaite}}, \bibinfo {author}
  {\bibfnamefont {L.}~\bibnamefont {Gaudreau}}, \bibinfo {author}
  {\bibfnamefont {F.~S.}\ \bibnamefont {Cataliotti}}, \bibinfo {author}
  {\bibfnamefont {F.}~\bibnamefont {Koppens}}, \ and\ \bibinfo {author}
  {\bibfnamefont {C.}~\bibnamefont {Toninelli}},\ }\bibfield  {title}
  {{\selectlanguage {english}\enquote {\bibinfo {title} {Single-molecule study
  for a graphene-based nano-position sensor},}\ }}\href {\doibase
  10.1088/1367-2630/16/11/113007} {\bibfield  {journal} {\bibinfo  {journal}
  {New J. Phys.}\ }\textbf {\bibinfo {volume} {16}},\ \bibinfo {pages} {113007}
  (\bibinfo {year} {2014})}\BibitemShut {NoStop}%
\bibitem [{\citenamefont {Kucsko}\ \emph {et~al.}(2013)\citenamefont {Kucsko},
  \citenamefont {Maurer}, \citenamefont {Yao}, \citenamefont {Kubo},
  \citenamefont {Noh}, \citenamefont {Lo}, \citenamefont {Park},\ and\
  \citenamefont {Lukin}}]{kucsko_nanometre-scale_2013}%
  \BibitemOpen
  \bibfield  {author} {\bibinfo {author} {\bibfnamefont {G.}~\bibnamefont
  {Kucsko}}, \bibinfo {author} {\bibfnamefont {P.~C.}\ \bibnamefont {Maurer}},
  \bibinfo {author} {\bibfnamefont {N.~Y.}\ \bibnamefont {Yao}}, \bibinfo
  {author} {\bibfnamefont {M.}~\bibnamefont {Kubo}}, \bibinfo {author}
  {\bibfnamefont {H.~J.}\ \bibnamefont {Noh}}, \bibinfo {author} {\bibfnamefont
  {P.~K.}\ \bibnamefont {Lo}}, \bibinfo {author} {\bibfnamefont
  {H.}~\bibnamefont {Park}}, \ and\ \bibinfo {author} {\bibfnamefont {M.~D.}\
  \bibnamefont {Lukin}},\ }\bibfield  {title} {{\selectlanguage
  {english}\enquote {\bibinfo {title} {Nanometre-scale thermometry in a living
  cell},}\ }}\href {\doibase 10.1038/nature12373} {\bibfield  {journal}
  {\bibinfo  {journal} {Nature}\ }\textbf {\bibinfo {volume} {500}},\ \bibinfo
  {pages} {54--58} (\bibinfo {year} {2013})}\BibitemShut {NoStop}%
\bibitem [{\citenamefont {Lukin}\ and\ \citenamefont
  {Imamo\u{g}lu}(2001)}]{lukin_controlling_2001}%
  \BibitemOpen
  \bibfield  {author} {\bibinfo {author} {\bibfnamefont {M.~D.}\ \bibnamefont
  {Lukin}}\ and\ \bibinfo {author} {\bibfnamefont {A.}~\bibnamefont
  {Imamo\u{g}lu}},\ }\bibfield  {title} {{\selectlanguage {english}\enquote
  {\bibinfo {title} {Controlling photons using electromagnetically induced
  transparency},}\ }}\href {\doibase 10.1038/35095000} {\bibfield  {journal}
  {\bibinfo  {journal} {Nature}\ }\textbf {\bibinfo {volume} {413}},\ \bibinfo
  {pages} {273--276} (\bibinfo {year} {2001})}\BibitemShut {NoStop}%
\bibitem [{\citenamefont {Reim}\ \emph {et~al.}(2010)\citenamefont {Reim},
  \citenamefont {Nunn}, \citenamefont {Lorenz}, \citenamefont {Sussman},
  \citenamefont {Lee}, \citenamefont {Langford}, \citenamefont {Jaksch},\ and\
  \citenamefont {Walmsley}}]{reim_towards_2010}%
  \BibitemOpen
  \bibfield  {author} {\bibinfo {author} {\bibfnamefont {K.~F.}\ \bibnamefont
  {Reim}}, \bibinfo {author} {\bibfnamefont {J.}~\bibnamefont {Nunn}}, \bibinfo
  {author} {\bibfnamefont {V.~O.}\ \bibnamefont {Lorenz}}, \bibinfo {author}
  {\bibfnamefont {B.~J.}\ \bibnamefont {Sussman}}, \bibinfo {author}
  {\bibfnamefont {K.~C.}\ \bibnamefont {Lee}}, \bibinfo {author} {\bibfnamefont
  {N.~K.}\ \bibnamefont {Langford}}, \bibinfo {author} {\bibfnamefont
  {D.}~\bibnamefont {Jaksch}}, \ and\ \bibinfo {author} {\bibfnamefont {I.~A.}\
  \bibnamefont {Walmsley}},\ }\bibfield  {title} {{\selectlanguage
  {english}\enquote {\bibinfo {title} {Towards high-speed optical quantum
  memories},}\ }}\href {\doibase 10.1038/nphoton.2010.30} {\bibfield  {journal}
  {\bibinfo  {journal} {Nat. Photon.}\ }\textbf {\bibinfo {volume} {4}},\
  \bibinfo {pages} {218--221} (\bibinfo {year} {2010})}\BibitemShut {NoStop}%
\bibitem [{\citenamefont {Volz}\ \emph {et~al.}(2014)\citenamefont {Volz},
  \citenamefont {Scheucher}, \citenamefont {Junge},\ and\ \citenamefont
  {Rauschenbeutel}}]{volz_nonlinear_2014}%
  \BibitemOpen
  \bibfield  {author} {\bibinfo {author} {\bibfnamefont {J.}~\bibnamefont
  {Volz}}, \bibinfo {author} {\bibfnamefont {M.}~\bibnamefont {Scheucher}},
  \bibinfo {author} {\bibfnamefont {C.}~\bibnamefont {Junge}}, \ and\ \bibinfo
  {author} {\bibfnamefont {A.}~\bibnamefont {Rauschenbeutel}},\ }\bibfield
  {title} {{\selectlanguage {english}\enquote {\bibinfo {title} {Nonlinear π
  phase shift for single fibre-guided photons interacting with a single
  resonator-enhanced atom},}\ }}\href {\doibase 10.1038/nphoton.2014.253}
  {\bibfield  {journal} {\bibinfo  {journal} {Nat. Photon.}\ }\textbf {\bibinfo
  {volume} {8}},\ \bibinfo {pages} {965--970} (\bibinfo {year}
  {2014})}\BibitemShut {NoStop}%
\bibitem [{\citenamefont {Thompson}\ \emph {et~al.}(2013)\citenamefont
  {Thompson}, \citenamefont {Tiecke}, \citenamefont {de~Leon}, \citenamefont
  {Feist}, \citenamefont {Akimov}, \citenamefont {Gullans}, \citenamefont
  {Zibrov}, \citenamefont {Vuletic},\ and\ \citenamefont
  {Lukin}}]{thompson_coupling_2013}%
  \BibitemOpen
  \bibfield  {author} {\bibinfo {author} {\bibfnamefont {J.~D.}\ \bibnamefont
  {Thompson}}, \bibinfo {author} {\bibfnamefont {T.~G.}\ \bibnamefont
  {Tiecke}}, \bibinfo {author} {\bibfnamefont {N.~P.}\ \bibnamefont {de~Leon}},
  \bibinfo {author} {\bibfnamefont {J.}~\bibnamefont {Feist}}, \bibinfo
  {author} {\bibfnamefont {A.~V.}\ \bibnamefont {Akimov}}, \bibinfo {author}
  {\bibfnamefont {M.}~\bibnamefont {Gullans}}, \bibinfo {author} {\bibfnamefont
  {A.~S.}\ \bibnamefont {Zibrov}}, \bibinfo {author} {\bibfnamefont
  {V.}~\bibnamefont {Vuletic}, \bibfnamefont {V.}}, \ and\ \bibinfo {author}
  {\bibfnamefont {M.~D.}\ \bibnamefont {Lukin}},\ }\bibfield  {title} {\enquote
  {\bibinfo {title} {Coupling a {Single} {Trapped} {Atom} to a {Nanoscale}
  {Optical} {Cavity}},}\ }\href {\doibase 10.1126/science.1237125} {\bibfield
  {journal} {\bibinfo  {journal} {Science}\ }\textbf {\bibinfo {volume}
  {340}},\ \bibinfo {pages} {1202--1205} (\bibinfo {year} {2013})}\BibitemShut
  {NoStop}%
\bibitem [{\citenamefont {Ritter}\ \emph {et~al.}(2012)\citenamefont {Ritter},
  \citenamefont {N\"{o}lleke}, \citenamefont {Hahn}, \citenamefont {Reiserer},
  \citenamefont {Neuzner}, \citenamefont {Uphoff}, \citenamefont {M\"{u}cke},
  \citenamefont {Figueroa}, \citenamefont {Bochmann},\ and\ \citenamefont
  {Rempe}}]{ritter_elementary_2012}%
  \BibitemOpen
  \bibfield  {author} {\bibinfo {author} {\bibfnamefont {S.}~\bibnamefont
  {Ritter}}, \bibinfo {author} {\bibfnamefont {C.}~\bibnamefont {N\"{o}lleke}},
  \bibinfo {author} {\bibfnamefont {C.}~\bibnamefont {Hahn}}, \bibinfo {author}
  {\bibfnamefont {A.}~\bibnamefont {Reiserer}}, \bibinfo {author}
  {\bibfnamefont {A.}~\bibnamefont {Neuzner}}, \bibinfo {author} {\bibfnamefont
  {M.}~\bibnamefont {Uphoff}}, \bibinfo {author} {\bibfnamefont
  {M.}~\bibnamefont {M\"{u}cke}}, \bibinfo {author} {\bibfnamefont
  {E.}~\bibnamefont {Figueroa}}, \bibinfo {author} {\bibfnamefont
  {J.}~\bibnamefont {Bochmann}}, \ and\ \bibinfo {author} {\bibfnamefont
  {G.}~\bibnamefont {Rempe}},\ }\bibfield  {title} {{\selectlanguage
  {english}\enquote {\bibinfo {title} {An elementary quantum network of single
  atoms in optical cavities},}\ }}\href {\doibase 10.1038/nature11023}
  {\bibfield  {journal} {\bibinfo  {journal} {Nature}\ }\textbf {\bibinfo
  {volume} {484}},\ \bibinfo {pages} {195--200} (\bibinfo {year}
  {2012})}\BibitemShut {NoStop}%
\bibitem [{\citenamefont {Schietinger}\ \emph {et~al.}(2008)\citenamefont
  {Schietinger}, \citenamefont {Schr\"{o}der},\ and\ \citenamefont
  {Benson}}]{schietinger_one-by-one_2008}%
  \BibitemOpen
  \bibfield  {author} {\bibinfo {author} {\bibfnamefont {S.}~\bibnamefont
  {Schietinger}}, \bibinfo {author} {\bibfnamefont {T.}~\bibnamefont
  {Schr\"{o}der}}, \ and\ \bibinfo {author} {\bibfnamefont {O.}~\bibnamefont
  {Benson}},\ }\bibfield  {title} {\enquote {\bibinfo {title} {One-by-{One}
  {Coupling} of {Single} {Defect} {Centers} in {Nanodiamonds} to {High}-{Q}
  {Modes} of an {Optical} {Microresonator}},}\ }\href {\doibase
  10.1021/nl8023627} {\bibfield  {journal} {\bibinfo  {journal} {Nano Lett.}\
  }\textbf {\bibinfo {volume} {8}},\ \bibinfo {pages} {3911--3915} (\bibinfo
  {year} {2008})}\BibitemShut {NoStop}%
\bibitem [{\citenamefont {Faez}\ \emph
  {et~al.}(2014{\natexlab{b}})\citenamefont {Faez}, \citenamefont
  {T\"{u}rschmann}, \citenamefont {Haakh}, \citenamefont {G\"{o}tzinger},\ and\
  \citenamefont {Sandoghdar}}]{faez_coherent_2014}%
  \BibitemOpen
  \bibfield  {author} {\bibinfo {author} {\bibfnamefont {S.}~\bibnamefont
  {Faez}}, \bibinfo {author} {\bibfnamefont {P.}~\bibnamefont
  {T\"{u}rschmann}}, \bibinfo {author} {\bibfnamefont {H.~R.}\ \bibnamefont
  {Haakh}}, \bibinfo {author} {\bibfnamefont {S.}~\bibnamefont
  {G\"{o}tzinger}}, \ and\ \bibinfo {author} {\bibfnamefont {V.}~\bibnamefont
  {Sandoghdar}},\ }\bibfield  {title} {\enquote {\bibinfo {title} {Coherent
  {Interaction} of {Light} and {Single} {Molecules} in a {Dielectric}
  {Nanoguide}},}\ }\href {\doibase 10.1103/PhysRevLett.113.213601} {\bibfield
  {journal} {\bibinfo  {journal} {Phys. Rev. Lett.}\ }\textbf {\bibinfo
  {volume} {113}},\ \bibinfo {pages} {213601} (\bibinfo {year}
  {2014}{\natexlab{b}})}\BibitemShut {NoStop}%
\bibitem [{\citenamefont {Hwang}\ and\ \citenamefont
  {Hinds}(2011)}]{hwang_dye_2011}%
  \BibitemOpen
  \bibfield  {author} {\bibinfo {author} {\bibfnamefont {J.}~\bibnamefont
  {Hwang}}\ and\ \bibinfo {author} {\bibfnamefont {E.~A.}\ \bibnamefont
  {Hinds}},\ }\bibfield  {title} {{\selectlanguage {english}\enquote {\bibinfo
  {title} {Dye molecules as single-photon sources and large optical
  nonlinearities on a chip},}\ }}\href {\doibase 10.1088/1367-2630/13/8/085009}
  {\bibfield  {journal} {\bibinfo  {journal} {New J. Phys.}\ }\textbf {\bibinfo
  {volume} {13}},\ \bibinfo {pages} {085009} (\bibinfo {year}
  {2011})}\BibitemShut {NoStop}%
\bibitem [{\citenamefont {Yalla}\ \emph {et~al.}(2012)\citenamefont {Yalla},
  \citenamefont {Le~Kien}, \citenamefont {Morinaga},\ and\ \citenamefont
  {Hakuta}}]{yalla_efficient_2012}%
  \BibitemOpen
  \bibfield  {author} {\bibinfo {author} {\bibfnamefont {R.}~\bibnamefont
  {Yalla}}, \bibinfo {author} {\bibfnamefont {Fam}\ \bibnamefont {Le~Kien}},
  \bibinfo {author} {\bibfnamefont {M.}~\bibnamefont {Morinaga}}, \ and\
  \bibinfo {author} {\bibfnamefont {K.}~\bibnamefont {Hakuta}},\ }\bibfield
  {title} {\enquote {\bibinfo {title} {Efficient {Channeling} of {Fluorescence}
  {Photons} from {Single} {Quantum} {Dots} into {Guided} {Modes} of {Optical}
  {Nanofiber}},}\ }\href {\doibase 10.1103/PhysRevLett.109.063602} {\bibfield
  {journal} {\bibinfo  {journal} {Phys. Rev. Lett.}\ }\textbf {\bibinfo
  {volume} {109}},\ \bibinfo {pages} {063602} (\bibinfo {year}
  {2012})}\BibitemShut {NoStop}%
\bibitem [{\citenamefont {Vetsch}\ \emph {et~al.}(2010)\citenamefont {Vetsch},
  \citenamefont {Reitz}, \citenamefont {Sagu\'{e}}, \citenamefont {Schmidt},
  \citenamefont {Dawkins},\ and\ \citenamefont
  {Rauschenbeutel}}]{vetsch_optical_2010}%
  \BibitemOpen
  \bibfield  {author} {\bibinfo {author} {\bibfnamefont {E.}~\bibnamefont
  {Vetsch}}, \bibinfo {author} {\bibfnamefont {D.}~\bibnamefont {Reitz}},
  \bibinfo {author} {\bibfnamefont {G.}~\bibnamefont {Sagu\'{e}}}, \bibinfo
  {author} {\bibfnamefont {R.}~\bibnamefont {Schmidt}}, \bibinfo {author}
  {\bibfnamefont {S.~T.}\ \bibnamefont {Dawkins}}, \ and\ \bibinfo {author}
  {\bibfnamefont {A.}~\bibnamefont {Rauschenbeutel}},\ }\bibfield  {title}
  {\enquote {\bibinfo {title} {Optical {Interface} {Created} by
  {Laser}-{Cooled} {Atoms} {Trapped} in the {Evanescent} {Field} {Surrounding}
  an {Optical} {Nanofiber}},}\ }\href {\doibase 10.1103/PhysRevLett.104.203603}
  {\bibfield  {journal} {\bibinfo  {journal} {Phys. Rev. Lett.}\ }\textbf
  {\bibinfo {volume} {104}},\ \bibinfo {pages} {203603} (\bibinfo {year}
  {2010})}\BibitemShut {NoStop}%
\bibitem [{\citenamefont {Kir\u{s}ansk\.{e}}\ \emph {et~al.}(2017)\citenamefont
  {Kir\u{s}ansk\.{e}}, \citenamefont {Thyrrestrup}, \citenamefont {Daveau},
  \citenamefont {Dreeßen}, \citenamefont {Pregnolato}, \citenamefont {Midolo},
  \citenamefont {Tighineanu}, \citenamefont {Javadi}, \citenamefont {Stobbe},
  \citenamefont {Schott}, \citenamefont {Ludwig}, \citenamefont {Wieck},
  \citenamefont {Park}, \citenamefont {Song}, \citenamefont {Kuhlmann},
  \citenamefont {S\"{o}llner}, \citenamefont {L\"{o}bl}, \citenamefont
  {Warburton},\ and\ \citenamefont {Lodahl}}]{kirsanske_qd_2017}%
  \BibitemOpen
  \bibfield  {author} {\bibinfo {author} {\bibfnamefont {G.}~\bibnamefont
  {Kir\u{s}ansk\.{e}}}, \bibinfo {author} {\bibfnamefont {H.}~\bibnamefont
  {Thyrrestrup}}, \bibinfo {author} {\bibfnamefont {R.~S.}\ \bibnamefont
  {Daveau}}, \bibinfo {author} {\bibfnamefont {C.~L.}\ \bibnamefont {Dreeßen}},
  \bibinfo {author} {\bibfnamefont {T.}~\bibnamefont {Pregnolato}}, \bibinfo
  {author} {\bibfnamefont {L.}~\bibnamefont {Midolo}}, \bibinfo {author}
  {\bibfnamefont {P.}~\bibnamefont {Tighineanu}}, \bibinfo {author}
  {\bibfnamefont {A.}~\bibnamefont {Javadi}}, \bibinfo {author} {\bibfnamefont
  {S.}~\bibnamefont {Stobbe}}, \bibinfo {author} {\bibfnamefont
  {R.}~\bibnamefont {Schott}}, \bibinfo {author} {\bibfnamefont
  {A.}~\bibnamefont {Ludwig}}, \bibinfo {author} {\bibfnamefont {A.~D.}\
  \bibnamefont {Wieck}}, \bibinfo {author} {\bibfnamefont {S.~I.}\ \bibnamefont
  {Park}}, \bibinfo {author} {\bibfnamefont {J.~D.}\ \bibnamefont {Song}},
  \bibinfo {author} {\bibfnamefont {A.~V.}\ \bibnamefont {Kuhlmann}}, \bibinfo
  {author} {\bibfnamefont {I.}~\bibnamefont {S\"{o}llner}}, \bibinfo {author}
  {\bibfnamefont {C.}~\bibnamefont {L\"{o}bl}}, \bibinfo {author}
  {\bibfnamefont {R.~J.}\ \bibnamefont {Warburton}}, \ and\ \bibinfo {author}
  {\bibfnamefont {P.}~\bibnamefont {Lodahl}},\ }\bibfield  {title} {\enquote
  {\bibinfo {title} {Indistinguishable and efficient single photons from a
  quantum dot in a planar nanobeam waveguide},}\ }\href@noop {} {\bibfield
  {journal} {\bibinfo  {journal} {arXiv1701.08131}\ } (\bibinfo {year}
  {2017})}\BibitemShut {NoStop}%
\bibitem [{\citenamefont {Lombardi}\ \emph {et~al.}()\citenamefont {Lombardi},
  \citenamefont {Ovvyan}, \citenamefont {Pazzagli}, \citenamefont {Mazzamuto},
  \citenamefont {Kewes}, \citenamefont {Neitzke}, \citenamefont {Gruhler},
  \citenamefont {Benson}, \citenamefont {Pernice}, \citenamefont {Cataliotti},\
  and\ \citenamefont {Toninelli}}]{lombardi_molecules_2017}%
  \BibitemOpen
  \bibfield  {author} {\bibinfo {author} {\bibfnamefont {P.~E.}\ \bibnamefont
  {Lombardi}}, \bibinfo {author} {\bibfnamefont {A.~P.}\ \bibnamefont
  {Ovvyan}}, \bibinfo {author} {\bibfnamefont {S.}~\bibnamefont {Pazzagli}},
  \bibinfo {author} {\bibfnamefont {G.}~\bibnamefont {Mazzamuto}}, \bibinfo
  {author} {\bibfnamefont {G.}~\bibnamefont {Kewes}}, \bibinfo {author}
  {\bibfnamefont {O.}~\bibnamefont {Neitzke}}, \bibinfo {author} {\bibfnamefont
  {N.}~\bibnamefont {Gruhler}}, \bibinfo {author} {\bibfnamefont
  {O.}~\bibnamefont {Benson}}, \bibinfo {author} {\bibfnamefont {W.~H.~P.}\
  \bibnamefont {Pernice}}, \bibinfo {author} {\bibfnamefont {F.~S.}\
  \bibnamefont {Cataliotti}}, \ and\ \bibinfo {author} {\bibfnamefont
  {C.}~\bibnamefont {Toninelli}},\ }\bibfield  {title} {\enquote {\bibinfo
  {title} {Photostable molecules on chip: Integrated sources of nonclassical
  light},}\ }\href@noop {} {\bibinfo  {journal} {ACS Photonics,
  doi:10.1021/acsphotonics.7b00521}\ }\BibitemShut {NoStop}%
\bibitem [{\citenamefont {Garcia-Fernandez}\ \emph {et~al.}(2011)\citenamefont
  {Garcia-Fernandez}, \citenamefont {Alt}, \citenamefont {Bruse}, \citenamefont
  {Dan}, \citenamefont {Karapetyan}, \citenamefont {Rehband}, \citenamefont
  {Stiebeiner}, \citenamefont {Wiedemann}, \citenamefont {Meschede},\ and\
  \citenamefont {Rauschenbeutel}}]{garcia-fernandez_optical_2011}%
  \BibitemOpen
\bibfield  {journal} {  }\bibfield  {author} {\bibinfo {author} {\bibfnamefont
  {R.}~\bibnamefont {Garcia-Fernandez}}, \bibinfo {author} {\bibfnamefont
  {W.}~\bibnamefont {Alt}}, \bibinfo {author} {\bibfnamefont {F.}~\bibnamefont
  {Bruse}}, \bibinfo {author} {\bibfnamefont {C.}~\bibnamefont {Dan}}, \bibinfo
  {author} {\bibfnamefont {K.}~\bibnamefont {Karapetyan}}, \bibinfo {author}
  {\bibfnamefont {O.}~\bibnamefont {Rehband}}, \bibinfo {author} {\bibfnamefont
  {A.}~\bibnamefont {Stiebeiner}}, \bibinfo {author} {\bibfnamefont
  {U.}~\bibnamefont {Wiedemann}}, \bibinfo {author} {\bibfnamefont
  {D.}~\bibnamefont {Meschede}}, \ and\ \bibinfo {author} {\bibfnamefont
  {A.}~\bibnamefont {Rauschenbeutel}},\ }\bibfield  {title} {\enquote {\bibinfo
  {title} {Optical nanofibers and spectroscopy},}\ }\href {\doibase
  10.1007/s00340-011-4730-x} {\bibfield  {journal} {\bibinfo  {journal} {Appl.
  Phys. B}\ }\textbf {\bibinfo {volume} {105}},\ \bibinfo {pages} {3--15}
  (\bibinfo {year} {2011})}\BibitemShut {NoStop}%
\bibitem [{\citenamefont {Le~Kien}\ \emph
  {et~al.}(2005{\natexlab{a}})\citenamefont {Le~Kien}, \citenamefont
  {Dutta~Gupta}, \citenamefont {Balykin},\ and\ \citenamefont
  {Hakuta}}]{le_kien_spontaneous_2005}%
  \BibitemOpen
  \bibfield  {author} {\bibinfo {author} {\bibfnamefont {Fam}\ \bibnamefont
  {Le~Kien}}, \bibinfo {author} {\bibfnamefont {S.}~\bibnamefont
  {Dutta~Gupta}}, \bibinfo {author} {\bibfnamefont {V.~I.}\ \bibnamefont
  {Balykin}}, \ and\ \bibinfo {author} {\bibfnamefont {K.}~\bibnamefont
  {Hakuta}},\ }\bibfield  {title} {\enquote {\bibinfo {title} {Spontaneous
  emission of a cesium atom near a nanofiber: {Efficient} coupling of light to
  guided modes},}\ }\href {\doibase 10.1103/PhysRevA.72.032509} {\bibfield
  {journal} {\bibinfo  {journal} {Phys. Rev. A}\ }\textbf {\bibinfo {volume}
  {72}},\ \bibinfo {pages} {032509} (\bibinfo {year}
  {2005}{\natexlab{a}})}\BibitemShut {NoStop}%
\bibitem [{\citenamefont {Reitz}\ \emph {et~al.}(2013)\citenamefont {Reitz},
  \citenamefont {Sayrin}, \citenamefont {Mitsch}, \citenamefont {Schneeweiss},\
  and\ \citenamefont {Rauschenbeutel}}]{reitz_coherence_2013}%
  \BibitemOpen
  \bibfield  {author} {\bibinfo {author} {\bibfnamefont {D.}~\bibnamefont
  {Reitz}}, \bibinfo {author} {\bibfnamefont {C.}~\bibnamefont {Sayrin}},
  \bibinfo {author} {\bibfnamefont {R.}~\bibnamefont {Mitsch}}, \bibinfo
  {author} {\bibfnamefont {P.}~\bibnamefont {Schneeweiss}}, \ and\ \bibinfo
  {author} {\bibfnamefont {A.}~\bibnamefont {Rauschenbeutel}},\ }\bibfield
  {title} {\enquote {\bibinfo {title} {Coherence {Properties} of
  {Nanofiber}-{Trapped} {Cesium} {Atoms}},}\ }\href {\doibase
  10.1103/PhysRevLett.110.243603} {\bibfield  {journal} {\bibinfo  {journal}
  {Phys. Rev. Lett.}\ }\textbf {\bibinfo {volume} {110}},\ \bibinfo {pages}
  {243603} (\bibinfo {year} {2013})}\BibitemShut {NoStop}%
\bibitem [{\citenamefont {Nayak}\ \emph {et~al.}(2007)\citenamefont {Nayak},
  \citenamefont {Melentiev}, \citenamefont {Morinaga}, \citenamefont {Kien},
  \citenamefont {Balykin},\ and\ \citenamefont {Hakuta}}]{nayak_optical_2007}%
  \BibitemOpen
  \bibfield  {author} {\bibinfo {author} {\bibfnamefont {K.~P.}\ \bibnamefont
  {Nayak}}, \bibinfo {author} {\bibfnamefont {P.~N.}\ \bibnamefont
  {Melentiev}}, \bibinfo {author} {\bibfnamefont {M.}~\bibnamefont {Morinaga}},
  \bibinfo {author} {\bibfnamefont {Fam~Le}\ \bibnamefont {Kien}}, \bibinfo
  {author} {\bibfnamefont {V.~I.}\ \bibnamefont {Balykin}}, \ and\ \bibinfo
  {author} {\bibfnamefont {K.}~\bibnamefont {Hakuta}},\ }\bibfield  {title}
  {\enquote {\bibinfo {title} {Optical nanofiber as an efficient tool for
  manipulating and probing {atomicFluorescence}},}\ }\href {\doibase
  10.1364/OE.15.005431} {\bibfield  {journal} {\bibinfo  {journal} {Opt.
  Express}\ }\textbf {\bibinfo {volume} {15}},\ \bibinfo {pages} {5431--5438}
  (\bibinfo {year} {2007})}\BibitemShut {NoStop}%
\bibitem [{\citenamefont {Liebermeister}\ \emph {et~al.}(2014)\citenamefont
  {Liebermeister}, \citenamefont {Petersen}, \citenamefont {M\"{u}nchow},
  \citenamefont {Burchardt}, \citenamefont {Hermelbracht}, \citenamefont
  {Tashima}, \citenamefont {Schell}, \citenamefont {Benson}, \citenamefont
  {Meinhardt}, \citenamefont {Krueger}, \citenamefont {Stiebeiner},
  \citenamefont {Rauschenbeutel}, \citenamefont {Weinfurter},\ and\
  \citenamefont {Weber}}]{liebermeister_tapered_2014}%
  \BibitemOpen
  \bibfield  {author} {\bibinfo {author} {\bibfnamefont {L.}~\bibnamefont
  {Liebermeister}}, \bibinfo {author} {\bibfnamefont {F.}~\bibnamefont
  {Petersen}}, \bibinfo {author} {\bibfnamefont {A.~v}\ \bibnamefont
  {M\"{u}nchow}}, \bibinfo {author} {\bibfnamefont {D.}~\bibnamefont
  {Burchardt}}, \bibinfo {author} {\bibfnamefont {J.}~\bibnamefont
  {Hermelbracht}}, \bibinfo {author} {\bibfnamefont {T.}~\bibnamefont
  {Tashima}}, \bibinfo {author} {\bibfnamefont {A.~W.}\ \bibnamefont {Schell}},
  \bibinfo {author} {\bibfnamefont {O.}~\bibnamefont {Benson}}, \bibinfo
  {author} {\bibfnamefont {T.}~\bibnamefont {Meinhardt}}, \bibinfo {author}
  {\bibfnamefont {A.}~\bibnamefont {Krueger}}, \bibinfo {author} {\bibfnamefont
  {A.}~\bibnamefont {Stiebeiner}}, \bibinfo {author} {\bibfnamefont
  {A.}~\bibnamefont {Rauschenbeutel}}, \bibinfo {author} {\bibfnamefont
  {H.}~\bibnamefont {Weinfurter}}, \ and\ \bibinfo {author} {\bibfnamefont
  {M.}~\bibnamefont {Weber}},\ }\bibfield  {title} {\enquote {\bibinfo {title}
  {Tapered fiber coupling of single photons emitted by a deterministically
  positioned single nitrogen vacancy center},}\ }\href {\doibase
  10.1063/1.4862207} {\bibfield  {journal} {\bibinfo  {journal} {Appl. Phys.
  Lett.}\ }\textbf {\bibinfo {volume} {104}},\ \bibinfo {pages} {031101}
  (\bibinfo {year} {2014})}\BibitemShut {NoStop}%
\bibitem [{\citenamefont {Kummer}\ \emph {et~al.}(1994)\citenamefont {Kummer},
  \citenamefont {Basch\'{e}},\ and\ \citenamefont
  {Br\"{a}uchle}}]{kummer_terrylene_1994}%
  \BibitemOpen
  \bibfield  {author} {\bibinfo {author} {\bibfnamefont {S.}~\bibnamefont
  {Kummer}}, \bibinfo {author} {\bibfnamefont {Th}~\bibnamefont {Basch\'{e}}},
  \ and\ \bibinfo {author} {\bibfnamefont {C.}~\bibnamefont {Br\"{a}uchle}},\
  }\bibfield  {title} {\enquote {\bibinfo {title} {Terrylene in p-terphenyl: a
  novel single crystalline system for single molecule spectroscopy at low
  temperatures},}\ }\href {\doibase
  http://dx.doi.org/10.1016/0009-2614(94)01043-9} {\bibfield  {journal}
  {\bibinfo  {journal} {Chem. Phys. Lett.}\ }\textbf {\bibinfo {volume}
  {229}},\ \bibinfo {pages} {309 -- 316} (\bibinfo {year} {1994})}\BibitemShut
  {NoStop}%
\bibitem [{\citenamefont {Moerner}(2004)}]{moerner_single-photon_2004}%
  \BibitemOpen
  \bibfield  {author} {\bibinfo {author} {\bibfnamefont {W.~E.}\ \bibnamefont
  {Moerner}},\ }\bibfield  {title} {\enquote {\bibinfo {title} {Single-photon
  sources based on single molecules in solids},}\ }\href
  {http://stacks.iop.org/1367-2630/6/i=1/a=088} {\bibfield  {journal} {\bibinfo
   {journal} {New J. Phys.}\ }\textbf {\bibinfo {volume} {6}},\ \bibinfo
  {pages} {88} (\bibinfo {year} {2004})}\BibitemShut {NoStop}%
\bibitem [{\citenamefont {Moerner}\ and\ \citenamefont
  {Orrit}(1999)}]{moerner_illuminating_1999}%
  \BibitemOpen
  \bibfield  {author} {\bibinfo {author} {\bibfnamefont {W.~E.}\ \bibnamefont
  {Moerner}}\ and\ \bibinfo {author} {\bibfnamefont {Michel}\ \bibnamefont
  {Orrit}},\ }\bibfield  {title} {\enquote {\bibinfo {title} {Illuminating
  {Single} {Molecules} in {Condensed} {Matter}},}\ }\href {\doibase
  10.1126/science.283.5408.1670} {\bibfield  {journal} {\bibinfo  {journal}
  {Science}\ }\textbf {\bibinfo {volume} {283}},\ \bibinfo {pages} {1670--1676}
  (\bibinfo {year} {1999})}\BibitemShut {NoStop}%
\bibitem [{\citenamefont {Michaelis}\ \emph {et~al.}(1999)\citenamefont
  {Michaelis}, \citenamefont {Hettich}, \citenamefont {Zayats}, \citenamefont
  {Eiermann}, \citenamefont {Mlynek},\ and\ \citenamefont
  {Sandoghdar}}]{michaelis_single_1999}%
  \BibitemOpen
  \bibfield  {author} {\bibinfo {author} {\bibfnamefont {J.}~\bibnamefont
  {Michaelis}}, \bibinfo {author} {\bibfnamefont {C.}~\bibnamefont {Hettich}},
  \bibinfo {author} {\bibfnamefont {A.}~\bibnamefont {Zayats}}, \bibinfo
  {author} {\bibfnamefont {B.}~\bibnamefont {Eiermann}}, \bibinfo {author}
  {\bibfnamefont {J.}~\bibnamefont {Mlynek}}, \ and\ \bibinfo {author}
  {\bibfnamefont {V.}~\bibnamefont {Sandoghdar}},\ }\bibfield  {title}
  {{\selectlanguage {english}\enquote {\bibinfo {title} {A single molecule as a
  probe of optical intensity distribution},}\ }}\href {\doibase
  10.1364/OL.24.000581} {\bibfield  {journal} {\bibinfo  {journal} {Opt.
  Lett.}\ }\textbf {\bibinfo {volume} {24}},\ \bibinfo {pages} {581} (\bibinfo
  {year} {1999})}\BibitemShut {NoStop}%
\bibitem [{\citenamefont {Tamarat}\ \emph {et~al.}(2000)\citenamefont
  {Tamarat}, \citenamefont {Maali}, \citenamefont {Lounis},\ and\ \citenamefont
  {Orrit}}]{tamarat_ten_2000}%
  \BibitemOpen
  \bibfield  {author} {\bibinfo {author} {\bibfnamefont {Ph.}\ \bibnamefont
  {Tamarat}}, \bibinfo {author} {\bibfnamefont {A.}~\bibnamefont {Maali}},
  \bibinfo {author} {\bibfnamefont {B.}~\bibnamefont {Lounis}}, \ and\ \bibinfo
  {author} {\bibfnamefont {M.}~\bibnamefont {Orrit}},\ }\bibfield  {title}
  {\enquote {\bibinfo {title} {Ten {Years} of {Single}-{Molecule}
  {Spectroscopy}},}\ }\href {\doibase 10.1021/jp992505l} {\bibfield  {journal}
  {\bibinfo  {journal} {J. Phys. Chem. A}\ }\textbf {\bibinfo {volume} {104}},\
  \bibinfo {pages} {1--16} (\bibinfo {year} {2000})}\BibitemShut {NoStop}%
\bibitem [{\citenamefont {T\"{u}rschmann}\ \emph {et~al.}(2017)\citenamefont
  {T\"{u}rschmann}, \citenamefont {Rotenberg}, \citenamefont {Renger},
  \citenamefont {Harder}, \citenamefont {Lohse}, \citenamefont {Utikal},
  \citenamefont {G\"{o}tzinger},\ and\ \citenamefont
  {Sandoghdar}}]{nanoguide_2017}%
  \BibitemOpen
  \bibfield  {author} {\bibinfo {author} {\bibfnamefont {P.}~\bibnamefont
  {T\"{u}rschmann}}, \bibinfo {author} {\bibfnamefont {N.}~\bibnamefont
  {Rotenberg}}, \bibinfo {author} {\bibfnamefont {J.}~\bibnamefont {Renger}},
  \bibinfo {author} {\bibfnamefont {I.}~\bibnamefont {Harder}}, \bibinfo
  {author} {\bibfnamefont {O.}~\bibnamefont {Lohse}}, \bibinfo {author}
  {\bibfnamefont {T.}~\bibnamefont {Utikal}}, \bibinfo {author} {\bibfnamefont
  {S.}~\bibnamefont {G\"{o}tzinger}}, \ and\ \bibinfo {author} {\bibfnamefont
  {V.}~\bibnamefont {Sandoghdar}},\ }\bibfield  {title} {\enquote {\bibinfo
  {title} {On-chip linear and nonlinear control of single molecules coupled to
  a nanoguide},}\ }\href@noop {} {\bibfield  {journal} {\bibinfo  {journal}
  {Nano Lett.}\ }\textbf {\bibinfo {volume} {17}},\ \bibinfo {pages}
  {4941--4945} (\bibinfo {year} {2017})}\BibitemShut {NoStop}%
\bibitem [{\citenamefont {Wang}\ \emph {et~al.}(2017)\citenamefont {Wang},
  \citenamefont {Kelkar}, \citenamefont {Martin-Cano}, \citenamefont {Utikal},
  \citenamefont {G\"{o}tzinger},\ and\ \citenamefont
  {Sandoghdar}}]{coherent_microcavity_2017}%
  \BibitemOpen
  \bibfield  {author} {\bibinfo {author} {\bibfnamefont {D.}~\bibnamefont
  {Wang}}, \bibinfo {author} {\bibfnamefont {H.}~\bibnamefont {Kelkar}},
  \bibinfo {author} {\bibfnamefont {D.}~\bibnamefont {Martin-Cano}}, \bibinfo
  {author} {\bibfnamefont {T.}~\bibnamefont {Utikal}}, \bibinfo {author}
  {\bibfnamefont {S.}~\bibnamefont {G\"{o}tzinger}}, \ and\ \bibinfo {author}
  {\bibfnamefont {V.}~\bibnamefont {Sandoghdar}},\ }\bibfield  {title}
  {\enquote {\bibinfo {title} {Coherent coupling of a single molecule to a
  scanning fabry-perot microcavity},}\ }\href@noop {} {\bibfield  {journal}
  {\bibinfo  {journal} {Physical Review X}\ }\textbf {\bibinfo {volume} {7}},\
  \bibinfo {pages} {021014} (\bibinfo {year} {2017})}\BibitemShut {NoStop}%
\bibitem [{\citenamefont {Chizhik}\ \emph {et~al.}(2009)\citenamefont
  {Chizhik}, \citenamefont {Schleifenbaum}, \citenamefont {Gutbrod},
  \citenamefont {Chizhik}, \citenamefont {Khoptyar},\ and\ \citenamefont
  {Meixner}}]{emission_microcavity_2009}%
  \BibitemOpen
  \bibfield  {author} {\bibinfo {author} {\bibfnamefont {A.}~\bibnamefont
  {Chizhik}}, \bibinfo {author} {\bibfnamefont {F.}~\bibnamefont
  {Schleifenbaum}}, \bibinfo {author} {\bibfnamefont {R.}~\bibnamefont
  {Gutbrod}}, \bibinfo {author} {\bibfnamefont {A.}~\bibnamefont {Chizhik}},
  \bibinfo {author} {\bibfnamefont {D.}~\bibnamefont {Khoptyar}}, \ and\
  \bibinfo {author} {\bibfnamefont {A.~J.}\ \bibnamefont {Meixner}},\
  }\bibfield  {title} {\enquote {\bibinfo {title} {Tuning the fluorescence
  emission spectra of a single molecule with a variable optical subwavelength
  metal microcavity},}\ }\href@noop {} {\bibfield  {journal} {\bibinfo
  {journal} {Phys. Rev. Lett.}\ }\textbf {\bibinfo {volume} {102}},\ \bibinfo
  {pages} {073002} (\bibinfo {year} {2009})}\BibitemShut {NoStop}%
\bibitem [{\citenamefont {Norris}\ \emph {et~al.}(1997)\citenamefont {Norris},
  \citenamefont {Kuwata-Gonokami},\ and\ \citenamefont
  {Moerner}}]{excitation_microcavity_1997}%
  \BibitemOpen
  \bibfield  {author} {\bibinfo {author} {\bibfnamefont {D.~J.}\ \bibnamefont
  {Norris}}, \bibinfo {author} {\bibfnamefont {M.}~\bibnamefont
  {Kuwata-Gonokami}}, \ and\ \bibinfo {author} {\bibfnamefont {W.~E.}\
  \bibnamefont {Moerner}},\ }\bibfield  {title} {\enquote {\bibinfo {title}
  {Excitation of a single molecule on the surface of a spherical
  microcavity},}\ }\href@noop {} {\bibfield  {journal} {\bibinfo  {journal}
  {Appl. Phys. Lett.}\ }\textbf {\bibinfo {volume} {71}} (\bibinfo {year}
  {1997})}\BibitemShut {NoStop}%
\bibitem [{\citenamefont {Warken}\ \emph {et~al.}(2008)\citenamefont {Warken},
  \citenamefont {Rauschenbeutel},\ and\ \citenamefont
  {Bartholomaus}}]{warken_fiber_2008}%
  \BibitemOpen
  \bibfield  {author} {\bibinfo {author} {\bibfnamefont {F.}~\bibnamefont
  {Warken}}, \bibinfo {author} {\bibfnamefont {A.}~\bibnamefont
  {Rauschenbeutel}}, \ and\ \bibinfo {author} {\bibfnamefont {T.}~\bibnamefont
  {Bartholomaus}},\ }\bibfield  {title} {\enquote {\bibinfo {title} {Fiber
  {Pulling} {Profits} from {Precise} {Positioning}-{Precise} motion control
  improves manufacturing of fiber optical resonators},}\ }\href@noop {}
  {\bibfield  {journal} {\bibinfo  {journal} {Photonics Spectra}\ }\textbf
  {\bibinfo {volume} {42}},\ \bibinfo {pages} {73} (\bibinfo {year}
  {2008})}\BibitemShut {NoStop}%
\bibitem [{\citenamefont {Stiebeiner}\ \emph {et~al.}(2010)\citenamefont
  {Stiebeiner}, \citenamefont {Garcia-Fernandez},\ and\ \citenamefont
  {Rauschenbeutel}}]{stiebeiner_design_2010}%
  \BibitemOpen
  \bibfield  {author} {\bibinfo {author} {\bibfnamefont {A.}~\bibnamefont
  {Stiebeiner}}, \bibinfo {author} {\bibfnamefont {R.}~\bibnamefont
  {Garcia-Fernandez}}, \ and\ \bibinfo {author} {\bibfnamefont
  {A.}~\bibnamefont {Rauschenbeutel}},\ }\bibfield  {title} {\enquote {\bibinfo
  {title} {Design and optimization of broadband tapered optical fibers with a
  nanofiber waist},}\ }\href {\doibase 10.1364/OE.18.022677} {\bibfield
  {journal} {\bibinfo  {journal} {Opt. Express}\ }\textbf {\bibinfo {volume}
  {18}},\ \bibinfo {pages} {22677--22685} (\bibinfo {year} {2010})}\BibitemShut
  {NoStop}%
\bibitem [{\citenamefont {Bordat}\ and\ \citenamefont
  {Brown}(2002)}]{bordat_molecular_2002}%
  \BibitemOpen
  \bibfield  {author} {\bibinfo {author} {\bibfnamefont {P.}~\bibnamefont
  {Bordat}}\ and\ \bibinfo {author} {\bibfnamefont {R.}~\bibnamefont {Brown}},\
  }\bibfield  {title} {{\selectlanguage {english}\enquote {\bibinfo {title}
  {Molecular mechanisms of photo-induced spectral diffusion of single terrylene
  molecules in p terphenyl},}\ }}\href {\doibase 10.1063/1.1420753} {\bibfield
  {journal} {\bibinfo  {journal} {J. Chem. Phys.}\ }\textbf {\bibinfo {volume}
  {116}},\ \bibinfo {pages} {229} (\bibinfo {year} {2002})}\BibitemShut
  {NoStop}%
\bibitem [{\citenamefont {Loudon}(2000)}]{loudon_quantum_2000}%
  \BibitemOpen
  \bibfield  {author} {\bibinfo {author} {\bibfnamefont {R.}~\bibnamefont
  {Loudon}},\ }\href@noop {} {{\selectlanguage {english}\emph {\bibinfo {title}
  {The {Quantum} {Theory} of {Light}}}}}\ (\bibinfo  {publisher} {OUP Oxford},\
  \bibinfo {year} {2000})\BibitemShut {NoStop}%
\bibitem [{\citenamefont {Warken}\ \emph {et~al.}(2007)\citenamefont {Warken},
  \citenamefont {Vetsch}, \citenamefont {Meschede}, \citenamefont
  {Sokolowski},\ and\ \citenamefont
  {Rauschenbeutel}}]{warken_ultra-sensitive_2007}%
  \BibitemOpen
  \bibfield  {author} {\bibinfo {author} {\bibfnamefont {F.}~\bibnamefont
  {Warken}}, \bibinfo {author} {\bibfnamefont {E.}~\bibnamefont {Vetsch}},
  \bibinfo {author} {\bibfnamefont {D.}~\bibnamefont {Meschede}}, \bibinfo
  {author} {\bibfnamefont {M.}~\bibnamefont {Sokolowski}}, \ and\ \bibinfo
  {author} {\bibfnamefont {A.}~\bibnamefont {Rauschenbeutel}},\ }\bibfield
  {title} {{\selectlanguage {english}\enquote {\bibinfo {title}
  {Ultra-sensitive surface absorption spectroscopy using sub-wavelength
  diameter optical fibers},}\ }}\href {\doibase 10.1364/OE.15.011952}
  {\bibfield  {journal} {\bibinfo  {journal} {Opt. Express}\ }\textbf {\bibinfo
  {volume} {15}},\ \bibinfo {pages} {11952} (\bibinfo {year}
  {2007})}\BibitemShut {NoStop}%
\bibitem [{\citenamefont {Banasiewicz}\ \emph {et~al.}(2005)\citenamefont
  {Banasiewicz}, \citenamefont {Morawski}, \citenamefont {Wiącek},\ and\
  \citenamefont {Kozankiewicz}}]{banasiewicz_triplet_2005}%
  \BibitemOpen
  \bibfield  {author} {\bibinfo {author} {\bibfnamefont {M.}~\bibnamefont
  {Banasiewicz}}, \bibinfo {author} {\bibfnamefont {O.}~\bibnamefont
  {Morawski}}, \bibinfo {author} {\bibfnamefont {D.}~\bibnamefont {Wiącek}}, \
  and\ \bibinfo {author} {\bibfnamefont {B.}~\bibnamefont {Kozankiewicz}},\
  }\bibfield  {title} {\enquote {\bibinfo {title} {Triplet population and
  depopulation rates of single terrylene molecules in p-terphenyl crystal},}\
  }\href {\doibase 10.1016/j.cplett.2005.08.120} {\bibfield  {journal}
  {\bibinfo  {journal} {Chem. Phys. Lett.}\ }\textbf {\bibinfo {volume}
  {414}},\ \bibinfo {pages} {374--377} (\bibinfo {year} {2005})}\BibitemShut
  {NoStop}%
\bibitem [{\citenamefont {Li}\ and\ \citenamefont
  {Egerton}(2004)}]{li_radiation_2004}%
  \BibitemOpen
  \bibfield  {author} {\bibinfo {author} {\bibfnamefont {P.}~\bibnamefont
  {Li}}\ and\ \bibinfo {author} {\bibfnamefont {R.~F.}\ \bibnamefont
  {Egerton}},\ }\bibfield  {title} {\enquote {\bibinfo {title} {Radiation
  damage in coronene, rubrene and p-terphenyl, measured for incident electrons
  of kinetic energy between 100 and 200 kev},}\ }\href {\doibase
  10.1016/j.ultramic.2004.05.010} {\bibfield  {journal} {\bibinfo  {journal}
  {Ultramicroscopy}\ }\textbf {\bibinfo {volume} {101}},\ \bibinfo {pages}
  {161--172} (\bibinfo {year} {2004})}\BibitemShut {NoStop}%
\bibitem [{\citenamefont {Rice}\ \emph {et~al.}(2012)\citenamefont {Rice},
  \citenamefont {Tham},\ and\ \citenamefont
  {Chronister}}]{rice_temperature_2012}%
  \BibitemOpen
  \bibfield  {author} {\bibinfo {author} {\bibfnamefont {Andrew~P.}\
  \bibnamefont {Rice}}, \bibinfo {author} {\bibfnamefont {Fook~S.}\
  \bibnamefont {Tham}}, \ and\ \bibinfo {author} {\bibfnamefont {Eric~L.}\
  \bibnamefont {Chronister}},\ }\bibfield  {title} {{\selectlanguage
  {english}\enquote {\bibinfo {title} {A {Temperature} {Dependent} {X}-ray
  {Study} of the {Order}–{Disorder} {Enantiotropic} {Phase} {Transition} of
  p-{Terphenyl}},}\ }}\href {\doibase 10.1007/s10870-012-0378-6} {\bibfield
  {journal} {\bibinfo  {journal} {J. Chem. Crystallogr.}\ }\textbf {\bibinfo
  {volume} {43}},\ \bibinfo {pages} {14--25} (\bibinfo {year}
  {2012})}\BibitemShut {NoStop}%
\bibitem [{\citenamefont {Kasai}\ \emph {et~al.}(1992)\citenamefont {Kasai},
  \citenamefont {Nalwa}, \citenamefont {Oikawa}, \citenamefont {Okada},
  \citenamefont {Matsuda}, \citenamefont {Minami}, \citenamefont {Kakuta},
  \citenamefont {Ono}, \citenamefont {Mukoh},\ and\ \citenamefont
  {Nakanishi}}]{kasai_novel_1992}%
  \BibitemOpen
  \bibfield  {author} {\bibinfo {author} {\bibfnamefont {H.}~\bibnamefont
  {Kasai}}, \bibinfo {author} {\bibfnamefont {H.~S.}\ \bibnamefont {Nalwa}},
  \bibinfo {author} {\bibfnamefont {H.}~\bibnamefont {Oikawa}}, \bibinfo
  {author} {\bibfnamefont {S.}~\bibnamefont {Okada}}, \bibinfo {author}
  {\bibfnamefont {H.}~\bibnamefont {Matsuda}}, \bibinfo {author} {\bibfnamefont
  {N.}~\bibnamefont {Minami}}, \bibinfo {author} {\bibfnamefont
  {A.}~\bibnamefont {Kakuta}}, \bibinfo {author} {\bibfnamefont
  {K.}~\bibnamefont {Ono}}, \bibinfo {author} {\bibfnamefont {A.}~\bibnamefont
  {Mukoh}}, \ and\ \bibinfo {author} {\bibfnamefont {H.}~\bibnamefont
  {Nakanishi}},\ }\bibfield  {title} {{\selectlanguage {english}\enquote
  {\bibinfo {title} {A {Novel} {Preparation} {Method} of {Organic}
  {Microcrystals}},}\ }}\href {\doibase 10.1143/JJAP.31.L1132} {\bibfield
  {journal} {\bibinfo  {journal} {Jpn. J. Appl. Phys.}\ }\textbf {\bibinfo
  {volume} {31}},\ \bibinfo {pages} {L1132--L1134} (\bibinfo {year}
  {1992})}\BibitemShut {NoStop}%
\bibitem [{\citenamefont {Dolde}\ \emph {et~al.}(2011)\citenamefont {Dolde},
  \citenamefont {Fedder}, \citenamefont {Doherty}, \citenamefont {N\"{o}bauer},
  \citenamefont {Rempp}, \citenamefont {Balasubramanian}, \citenamefont {Wolf},
  \citenamefont {Reinhard}, \citenamefont {Hollenberg}, \citenamefont
  {Jelezko},\ and\ \citenamefont {Wrachtrup}}]{dolde_electric-field_2011}%
  \BibitemOpen
  \bibfield  {author} {\bibinfo {author} {\bibfnamefont {F.}~\bibnamefont
  {Dolde}}, \bibinfo {author} {\bibfnamefont {H.}~\bibnamefont {Fedder}},
  \bibinfo {author} {\bibfnamefont {M.~W.}\ \bibnamefont {Doherty}}, \bibinfo
  {author} {\bibfnamefont {T.}~\bibnamefont {N\"{o}bauer}}, \bibinfo {author}
  {\bibfnamefont {F.}~\bibnamefont {Rempp}}, \bibinfo {author} {\bibfnamefont
  {G.}~\bibnamefont {Balasubramanian}}, \bibinfo {author} {\bibfnamefont
  {T.}~\bibnamefont {Wolf}}, \bibinfo {author} {\bibfnamefont {F.}~\bibnamefont
  {Reinhard}}, \bibinfo {author} {\bibfnamefont {L.~C.~L.}\ \bibnamefont
  {Hollenberg}}, \bibinfo {author} {\bibfnamefont {F.}~\bibnamefont {Jelezko}},
  \ and\ \bibinfo {author} {\bibfnamefont {J.}~\bibnamefont {Wrachtrup}},\
  }\bibfield  {title} {{\selectlanguage {english}\enquote {\bibinfo {title}
  {Electric-field sensing using single diamond spins},}\ }}\href {\doibase
  10.1038/nphys1969} {\bibfield  {journal} {\bibinfo  {journal} {Nat. Phys.}\
  }\textbf {\bibinfo {volume} {7}},\ \bibinfo {pages} {459--463} (\bibinfo
  {year} {2011})}\BibitemShut {NoStop}%
\bibitem [{\citenamefont {Robledo}\ \emph {et~al.}(2010)\citenamefont
  {Robledo}, \citenamefont {Bernien}, \citenamefont {van Weperen},\ and\
  \citenamefont {Hanson}}]{nvcenters_control_2010}%
  \BibitemOpen
  \bibfield  {author} {\bibinfo {author} {\bibfnamefont {Lucio}\ \bibnamefont
  {Robledo}}, \bibinfo {author} {\bibfnamefont {Hannes}\ \bibnamefont
  {Bernien}}, \bibinfo {author} {\bibfnamefont {Ilse}\ \bibnamefont {van
  Weperen}}, \ and\ \bibinfo {author} {\bibfnamefont {Ronald}\ \bibnamefont
  {Hanson}},\ }\bibfield  {title} {\enquote {\bibinfo {title} {Control and
  coherence of the optical transition of single nitrogen vacancy centers in
  diamond},}\ }\href@noop {} {\bibfield  {journal} {\bibinfo  {journal}
  {Physical Review Letters}\ }\textbf {\bibinfo {volume} {105}},\ \bibinfo
  {pages} {177403} (\bibinfo {year} {2010})}\BibitemShut {NoStop}%
\bibitem [{\citenamefont {Hansom}\ \emph
  {et~al.}(2014{\natexlab{b}})\citenamefont {Hansom}, \citenamefont {Schulte},
  \citenamefont {Matthiesen}, \citenamefont {Stanley},\ and\ \citenamefont
  {Atatüre}}]{qdot_stabilization_2014}%
  \BibitemOpen
  \bibfield  {author} {\bibinfo {author} {\bibfnamefont {Jack}\ \bibnamefont
  {Hansom}}, \bibinfo {author} {\bibfnamefont {Carsten H.~H.}\ \bibnamefont
  {Schulte}}, \bibinfo {author} {\bibfnamefont {Clemens}\ \bibnamefont
  {Matthiesen}}, \bibinfo {author} {\bibfnamefont {Megan~J.}\ \bibnamefont
  {Stanley}}, \ and\ \bibinfo {author} {\bibfnamefont {Mete}\ \bibnamefont
  {Atatüre}},\ }\bibfield  {title} {\enquote {\bibinfo {title} {Frequency
  stabilization of the zero-phonon line of a quantum dot via phonon-assisted
  active feedback},}\ }\href@noop {} {\bibfield  {journal} {\bibinfo  {journal}
  {Appl. Phys. Lett.}\ }\textbf {\bibinfo {volume} {105}},\ \bibinfo {pages}
  {172107} (\bibinfo {year} {2014}{\natexlab{b}})}\BibitemShut {NoStop}%
\bibitem [{\citenamefont {Verberk}\ and\ \citenamefont
  {Orrit}(2003)}]{verberk_photon_2003}%
  \BibitemOpen
  \bibfield  {author} {\bibinfo {author} {\bibfnamefont {R.}~\bibnamefont
  {Verberk}}\ and\ \bibinfo {author} {\bibfnamefont {M.}~\bibnamefont
  {Orrit}},\ }\bibfield  {title} {\enquote {\bibinfo {title} {Photon statistics
  in the fluorescence of single molecules and nanocrystals: {Correlation}
  functions versus distributions of on- and off-times},}\ }\href {\doibase
  10.1063/1.1582848} {\bibfield  {journal} {\bibinfo  {journal} {J. Chem.
  Phys.}\ }\textbf {\bibinfo {volume} {119}},\ \bibinfo {pages} {2214--2222}
  (\bibinfo {year} {2003})}\BibitemShut {NoStop}%
\bibitem [{\citenamefont {Tamarat}\ \emph {et~al.}(1995)\citenamefont
  {Tamarat}, \citenamefont {Lounis}, \citenamefont {Bernard}, \citenamefont
  {Orrit}, \citenamefont {Kummer}, \citenamefont {Kettner}, \citenamefont
  {Mais},\ and\ \citenamefont {Basch\'{e}}}]{tamarat_pump-probe_1995}%
  \BibitemOpen
  \bibfield  {author} {\bibinfo {author} {\bibfnamefont {Ph.}\ \bibnamefont
  {Tamarat}}, \bibinfo {author} {\bibfnamefont {B.}~\bibnamefont {Lounis}},
  \bibinfo {author} {\bibfnamefont {J.}~\bibnamefont {Bernard}}, \bibinfo
  {author} {\bibfnamefont {M.}~\bibnamefont {Orrit}}, \bibinfo {author}
  {\bibfnamefont {S.}~\bibnamefont {Kummer}}, \bibinfo {author} {\bibfnamefont
  {R.}~\bibnamefont {Kettner}}, \bibinfo {author} {\bibfnamefont
  {S.}~\bibnamefont {Mais}}, \ and\ \bibinfo {author} {\bibfnamefont {Th.}\
  \bibnamefont {Basch\'{e}}},\ }\bibfield  {title} {\enquote {\bibinfo {title}
  {Pump-{Probe} {Experiments} with a {Single} {Molecule}: ac-{Stark} {Effect}
  and {Nonlinear} {Optical} {Response}},}\ }\href {\doibase
  10.1103/PhysRevLett.75.1514} {\bibfield  {journal} {\bibinfo  {journal}
  {Phys. Rev. Lett.}\ }\textbf {\bibinfo {volume} {75}},\ \bibinfo {pages}
  {1514--1517} (\bibinfo {year} {1995})}\BibitemShut {NoStop}%
\bibitem [{\citenamefont {Fujiwara}\ \emph {et~al.}(2011)\citenamefont
  {Fujiwara}, \citenamefont {Toubaru}, \citenamefont {Noda}, \citenamefont
  {Zhao},\ and\ \citenamefont {Takeuchi}}]{fujiwara_highly_2011}%
  \BibitemOpen
  \bibfield  {author} {\bibinfo {author} {\bibfnamefont {M.}~\bibnamefont
  {Fujiwara}}, \bibinfo {author} {\bibfnamefont {K.}~\bibnamefont {Toubaru}},
  \bibinfo {author} {\bibfnamefont {T.}~\bibnamefont {Noda}}, \bibinfo {author}
  {\bibfnamefont {H.}~\bibnamefont {Zhao}}, \ and\ \bibinfo {author}
  {\bibfnamefont {S.}~\bibnamefont {Takeuchi}},\ }\bibfield  {title} {\enquote
  {\bibinfo {title} {Highly {Efficient} {Coupling} of {Photons} from
  {Nanoemitters} into {Single}-{Mode} {Optical} {Fibers}},}\ }\href {\doibase
  10.1021/nl2024867} {\bibfield  {journal} {\bibinfo  {journal} {Nano Lett.}\
  }\textbf {\bibinfo {volume} {11}},\ \bibinfo {pages} {4362--4365} (\bibinfo
  {year} {2011})}\BibitemShut {NoStop}%
\bibitem [{\citenamefont {Schr\"{o}der}\ \emph {et~al.}(2011)\citenamefont
  {Schr\"{o}der}, \citenamefont {G\"{a}deke}, \citenamefont {Banholzer},\ and\
  \citenamefont {Benson}}]{solid_immersion_2011}%
  \BibitemOpen
  \bibfield  {author} {\bibinfo {author} {\bibfnamefont {T.}~\bibnamefont
  {Schr\"{o}der}}, \bibinfo {author} {\bibfnamefont {F.}~\bibnamefont
  {G\"{a}deke}}, \bibinfo {author} {\bibfnamefont {M.~J.}\ \bibnamefont
  {Banholzer}}, \ and\ \bibinfo {author} {\bibfnamefont {O.}~\bibnamefont
  {Benson}},\ }\bibfield  {title} {\enquote {\bibinfo {title} {Ultrabright and
  efficient single-photon generation bassed on nitrogen-vacancy centres in
  nanodiamonds on a solid immersion lens},}\ }\href@noop {} {\bibfield
  {journal} {\bibinfo  {journal} {New J. Phys.}\ }\textbf {\bibinfo {volume}
  {13}},\ \bibinfo {pages} {055017} (\bibinfo {year} {2011})}\BibitemShut
  {NoStop}%
\bibitem [{\citenamefont {Karlsson}\ \emph {et~al.}(2016)\citenamefont
  {Karlsson}, \citenamefont {Rippe},\ and\ \citenamefont
  {Kr\"{o}ll}}]{karlsson_confocal_2016}%
  \BibitemOpen
  \bibfield  {author} {\bibinfo {author} {\bibfnamefont {J.}~\bibnamefont
  {Karlsson}}, \bibinfo {author} {\bibfnamefont {L.}~\bibnamefont {Rippe}}, \
  and\ \bibinfo {author} {\bibfnamefont {S.}~\bibnamefont {Kr\"{o}ll}},\
  }\bibfield  {title} {\enquote {\bibinfo {title} {A confocal optical
  microscope for detection of single impurities in a bulk crystal at cryogenic
  temperatures},}\ }\href@noop {} {\bibfield  {journal} {\bibinfo  {journal}
  {Rev. Sci. Instrum.}\ }\textbf {\bibinfo {volume} {87}},\ \bibinfo {pages}
  {033701} (\bibinfo {year} {2016})}\BibitemShut {NoStop}%
\bibitem [{\citenamefont {Goban}\ \emph {et~al.}(2012)\citenamefont {Goban},
  \citenamefont {Choi}, \citenamefont {Alton}, \citenamefont {Ding},
  \citenamefont {Lacroûte}, \citenamefont {Pototschnig}, \citenamefont
  {Thiele}, \citenamefont {Stern},\ and\ \citenamefont
  {Kimble}}]{goban_demonstration_2012}%
  \BibitemOpen
  \bibfield  {author} {\bibinfo {author} {\bibfnamefont {A.}~\bibnamefont
  {Goban}}, \bibinfo {author} {\bibfnamefont {K.~S.}\ \bibnamefont {Choi}},
  \bibinfo {author} {\bibfnamefont {D.~J.}\ \bibnamefont {Alton}}, \bibinfo
  {author} {\bibfnamefont {D.}~\bibnamefont {Ding}}, \bibinfo {author}
  {\bibfnamefont {C.}~\bibnamefont {Lacroûte}}, \bibinfo {author}
  {\bibfnamefont {M.}~\bibnamefont {Pototschnig}}, \bibinfo {author}
  {\bibfnamefont {T.}~\bibnamefont {Thiele}}, \bibinfo {author} {\bibfnamefont
  {N.~P.}\ \bibnamefont {Stern}}, \ and\ \bibinfo {author} {\bibfnamefont
  {H.~J.}\ \bibnamefont {Kimble}},\ }\bibfield  {title} {\enquote {\bibinfo
  {title} {Demonstration of a {State}-{Insensitive}, {Compensated} {Nanofiber}
  {Trap}},}\ }\href {\doibase 10.1103/PhysRevLett.109.033603} {\bibfield
  {journal} {\bibinfo  {journal} {Phys. Rev. Lett.}\ }\textbf {\bibinfo
  {volume} {109}},\ \bibinfo {pages} {033603} (\bibinfo {year}
  {2012})}\BibitemShut {NoStop}%
\bibitem [{\citenamefont {Hafezi}\ \emph {et~al.}(2012)\citenamefont {Hafezi},
  \citenamefont {Kim}, \citenamefont {Rolston}, \citenamefont {Orozco},
  \citenamefont {Lev},\ and\ \citenamefont {Taylor}}]{hafezi_atomic_2012}%
  \BibitemOpen
  \bibfield  {author} {\bibinfo {author} {\bibfnamefont {M.}~\bibnamefont
  {Hafezi}}, \bibinfo {author} {\bibfnamefont {Z.}~\bibnamefont {Kim}},
  \bibinfo {author} {\bibfnamefont {S.~L.}\ \bibnamefont {Rolston}}, \bibinfo
  {author} {\bibfnamefont {L.~A.}\ \bibnamefont {Orozco}}, \bibinfo {author}
  {\bibfnamefont {B.~L.}\ \bibnamefont {Lev}}, \ and\ \bibinfo {author}
  {\bibfnamefont {J.~M.}\ \bibnamefont {Taylor}},\ }\bibfield  {title}
  {\enquote {\bibinfo {title} {Atomic interface between microwave and optical
  photons},}\ }\href {\doibase 10.1103/PhysRevA.85.020302} {\bibfield
  {journal} {\bibinfo  {journal} {Phys. Rev. A}\ }\textbf {\bibinfo {volume}
  {85}},\ \bibinfo {pages} {020302} (\bibinfo {year} {2012})}\BibitemShut
  {NoStop}%
\bibitem [{\citenamefont {Fujiwara}\ \emph {et~al.}(2015)\citenamefont
  {Fujiwara}, \citenamefont {Zhao}, \citenamefont {Noda}, \citenamefont
  {Ikeda}, \citenamefont {Sumiya},\ and\ \citenamefont
  {Takeuchi}}]{fujiwara_ultrathin_2015}%
  \BibitemOpen
  \bibfield  {author} {\bibinfo {author} {\bibfnamefont {M.}~\bibnamefont
  {Fujiwara}}, \bibinfo {author} {\bibfnamefont {H.}~\bibnamefont {Zhao}},
  \bibinfo {author} {\bibfnamefont {T.}~\bibnamefont {Noda}}, \bibinfo {author}
  {\bibfnamefont {K.}~\bibnamefont {Ikeda}}, \bibinfo {author} {\bibfnamefont
  {H.}~\bibnamefont {Sumiya}}, \ and\ \bibinfo {author} {\bibfnamefont
  {S.}~\bibnamefont {Takeuchi}},\ }\bibfield  {title} {{\selectlanguage
  {english}\enquote {\bibinfo {title} {Ultrathin fiber-taper coupling with
  nitrogen vacancy centers in nanodiamonds at cryogenic temperatures},}\
  }}\href {\doibase 10.1364/OL.40.005702} {\bibfield  {journal} {\bibinfo
  {journal} {Opt. Lett.}\ }\textbf {\bibinfo {volume} {40}},\ \bibinfo {pages}
  {5702} (\bibinfo {year} {2015})}\BibitemShut {NoStop}%
\bibitem [{\citenamefont {Lounis}\ and\ \citenamefont
  {Moerner}(2000)}]{lounis_single_2000}%
  \BibitemOpen
  \bibfield  {author} {\bibinfo {author} {\bibfnamefont {B.}~\bibnamefont
  {Lounis}}\ and\ \bibinfo {author} {\bibfnamefont {W.~E.}\ \bibnamefont
  {Moerner}},\ }\bibfield  {title} {\enquote {\bibinfo {title} {Single photons
  on demand from a single molecule at room temperature},}\ }\href {\doibase
  10.1038/35035032} {\bibfield  {journal} {\bibinfo  {journal} {Nature}\
  }\textbf {\bibinfo {volume} {407}},\ \bibinfo {pages} {491--493} (\bibinfo
  {year} {2000})}\BibitemShut {NoStop}%
\bibitem [{\citenamefont {Hwang}\ \emph {et~al.}(2009)\citenamefont {Hwang},
  \citenamefont {Pototschnig}, \citenamefont {Lettow}, \citenamefont {Zumofen},
  \citenamefont {Renn}, \citenamefont {G\"{o}tzinger},\ and\ \citenamefont
  {Sandoghdar}}]{hwang_single-molecule_2009}%
  \BibitemOpen
  \bibfield  {author} {\bibinfo {author} {\bibfnamefont {J.}~\bibnamefont
  {Hwang}}, \bibinfo {author} {\bibfnamefont {M.}~\bibnamefont {Pototschnig}},
  \bibinfo {author} {\bibfnamefont {R.}~\bibnamefont {Lettow}}, \bibinfo
  {author} {\bibfnamefont {G.}~\bibnamefont {Zumofen}}, \bibinfo {author}
  {\bibfnamefont {A.}~\bibnamefont {Renn}}, \bibinfo {author} {\bibfnamefont
  {S.}~\bibnamefont {G\"{o}tzinger}}, \ and\ \bibinfo {author} {\bibfnamefont
  {V.}~\bibnamefont {Sandoghdar}},\ }\bibfield  {title} {\enquote {\bibinfo
  {title} {A single-molecule optical transistor},}\ }\href {\doibase
  10.1038/nature08134} {\bibfield  {journal} {\bibinfo  {journal} {Nature}\
  }\textbf {\bibinfo {volume} {460}},\ \bibinfo {pages} {76--80} (\bibinfo
  {year} {2009})}\BibitemShut {NoStop}%
\bibitem [{\citenamefont {Pototschnig}\ \emph {et~al.}(2011)\citenamefont
  {Pototschnig}, \citenamefont {Chassagneux}, \citenamefont {Hwang},
  \citenamefont {Zumofen}, \citenamefont {Renn},\ and\ \citenamefont
  {Sandoghdar}}]{pototschnig_controlling_2011}%
  \BibitemOpen
  \bibfield  {author} {\bibinfo {author} {\bibfnamefont {M.}~\bibnamefont
  {Pototschnig}}, \bibinfo {author} {\bibfnamefont {Y.}~\bibnamefont
  {Chassagneux}}, \bibinfo {author} {\bibfnamefont {J.}~\bibnamefont {Hwang}},
  \bibinfo {author} {\bibfnamefont {G.}~\bibnamefont {Zumofen}}, \bibinfo
  {author} {\bibfnamefont {A.}~\bibnamefont {Renn}}, \ and\ \bibinfo {author}
  {\bibfnamefont {V.}~\bibnamefont {Sandoghdar}},\ }\bibfield  {title}
  {\enquote {\bibinfo {title} {Controlling the {Phase} of a {Light} {Beam} with
  a {Single} {Molecule}},}\ }\href {\doibase 10.1103/PhysRevLett.107.063001}
  {\bibfield  {journal} {\bibinfo  {journal} {Phys. Rev. Lett.}\ }\textbf
  {\bibinfo {volume} {107}},\ \bibinfo {pages} {063001} (\bibinfo {year}
  {2011})}\BibitemShut {NoStop}%
\bibitem [{\citenamefont {Maser}\ \emph {et~al.}(2016)\citenamefont {Maser},
  \citenamefont {Gmeiner}, \citenamefont {Utikal}, \citenamefont
  {G\"{o}tzinger},\ and\ \citenamefont {Sandoghdar}}]{maser_few-photon_2016}%
  \BibitemOpen
  \bibfield  {author} {\bibinfo {author} {\bibfnamefont {A.}~\bibnamefont
  {Maser}}, \bibinfo {author} {\bibfnamefont {B.}~\bibnamefont {Gmeiner}},
  \bibinfo {author} {\bibfnamefont {T.}~\bibnamefont {Utikal}}, \bibinfo
  {author} {\bibfnamefont {S.}~\bibnamefont {G\"{o}tzinger}}, \ and\ \bibinfo
  {author} {\bibfnamefont {V.}~\bibnamefont {Sandoghdar}},\ }\bibfield  {title}
  {{\selectlanguage {english}\enquote {\bibinfo {title} {Few-photon coherent
  nonlinear optics with a single molecule},}\ }}\href {\doibase
  10.1038/nphoton.2016.63} {\bibfield  {journal} {\bibinfo  {journal} {Nat.
  Photon.}\ }\textbf {\bibinfo {volume} {10}},\ \bibinfo {pages} {450--453}
  (\bibinfo {year} {2016})}\BibitemShut {NoStop}%
\bibitem [{\citenamefont {Rist}\ \emph {et~al.}(2008)\citenamefont {Rist},
  \citenamefont {Eschner}, \citenamefont {Hennrich},\ and\ \citenamefont
  {Morigi}}]{rist_photon-mediated_2008}%
  \BibitemOpen
  \bibfield  {author} {\bibinfo {author} {\bibfnamefont {S.}~\bibnamefont
  {Rist}}, \bibinfo {author} {\bibfnamefont {J.}~\bibnamefont {Eschner}},
  \bibinfo {author} {\bibfnamefont {M.}~\bibnamefont {Hennrich}}, \ and\
  \bibinfo {author} {\bibfnamefont {G.}~\bibnamefont {Morigi}},\ }\bibfield
  {title} {\enquote {\bibinfo {title} {Photon-mediated interaction between two
  distant atoms},}\ }\href {\doibase 10.1103/PhysRevA.78.013808} {\bibfield
  {journal} {\bibinfo  {journal} {Phys. Rev. A}\ }\textbf {\bibinfo {volume}
  {78}},\ \bibinfo {pages} {013808} (\bibinfo {year} {2008})}\BibitemShut
  {NoStop}%
\bibitem [{\citenamefont {Loo}\ \emph {et~al.}(2013)\citenamefont {Loo},
  \citenamefont {Fedorov}, \citenamefont {Lalumière}, \citenamefont {Sanders},
  \citenamefont {Blais},\ and\ \citenamefont
  {Wallraff}}]{loo_photon-mediated_2013}%
  \BibitemOpen
  \bibfield  {author} {\bibinfo {author} {\bibfnamefont {A.~F.~van}\
  \bibnamefont {Loo}}, \bibinfo {author} {\bibfnamefont {A.}~\bibnamefont
  {Fedorov}}, \bibinfo {author} {\bibfnamefont {K.}~\bibnamefont {Lalumière}},
  \bibinfo {author} {\bibfnamefont {B.~C.}\ \bibnamefont {Sanders}}, \bibinfo
  {author} {\bibfnamefont {A.}~\bibnamefont {Blais}}, \ and\ \bibinfo {author}
  {\bibfnamefont {A.}~\bibnamefont {Wallraff}},\ }\bibfield  {title}
  {{\selectlanguage {english}\enquote {\bibinfo {title} {Photon-{Mediated}
  {Interactions} {Between} {Distant} {Artificial} {Atoms}},}\ }}\href {\doibase
  10.1126/science.1244324} {\bibfield  {journal} {\bibinfo  {journal}
  {Science}\ }\textbf {\bibinfo {volume} {342}},\ \bibinfo {pages} {1494--1496}
  (\bibinfo {year} {2013})}\BibitemShut {NoStop}%
\bibitem [{\citenamefont {Le~Kien}\ \emph
  {et~al.}(2005{\natexlab{b}})\citenamefont {Le~Kien}, \citenamefont {Gupta},
  \citenamefont {Nayak},\ and\ \citenamefont
  {Hakuta}}]{le_kien_nanofiber-mediated_2005}%
  \BibitemOpen
  \bibfield  {author} {\bibinfo {author} {\bibfnamefont {Fam}\ \bibnamefont
  {Le~Kien}}, \bibinfo {author} {\bibfnamefont {S.~Dutta}\ \bibnamefont
  {Gupta}}, \bibinfo {author} {\bibfnamefont {K.~P.}\ \bibnamefont {Nayak}}, \
  and\ \bibinfo {author} {\bibfnamefont {K.}~\bibnamefont {Hakuta}},\
  }\bibfield  {title} {\enquote {\bibinfo {title} {Nanofiber-mediated radiative
  transfer between two distant atoms},}\ }\href {\doibase
  10.1103/PhysRevA.72.063815} {\bibfield  {journal} {\bibinfo  {journal} {Phys.
  Rev. A}\ }\textbf {\bibinfo {volume} {72}},\ \bibinfo {pages} {063815}
  (\bibinfo {year} {2005}{\natexlab{b}})}\BibitemShut {NoStop}%
\bibitem [{\citenamefont {Wuttke}\ \emph {et~al.}(2012)\citenamefont {Wuttke},
  \citenamefont {Becker}, \citenamefont {Br\"{u}ckner}, \citenamefont
  {Rothhardt},\ and\ \citenamefont {Rauschenbeutel}}]{wuttke_nanofiber_2012}%
  \BibitemOpen
  \bibfield  {author} {\bibinfo {author} {\bibfnamefont {C.}~\bibnamefont
  {Wuttke}}, \bibinfo {author} {\bibfnamefont {M.}~\bibnamefont {Becker}},
  \bibinfo {author} {\bibfnamefont {S.}~\bibnamefont {Br\"{u}ckner}}, \bibinfo
  {author} {\bibfnamefont {M.}~\bibnamefont {Rothhardt}}, \ and\ \bibinfo
  {author} {\bibfnamefont {A.}~\bibnamefont {Rauschenbeutel}},\ }\bibfield
  {title} {\enquote {\bibinfo {title} {Nanofiber {Fabry}–{Perot}
  microresonator for nonlinear optics and cavity quantum electrodynamics},}\
  }\href {\doibase 10.1364/OL.37.001949} {\bibfield  {journal} {\bibinfo
  {journal} {Opt. Lett.}\ }\textbf {\bibinfo {volume} {37}},\ \bibinfo {pages}
  {1949--1951} (\bibinfo {year} {2012})}\BibitemShut {NoStop}%
\bibitem [{\citenamefont {Ahtee}\ \emph {et~al.}(2009)\citenamefont {Ahtee},
  \citenamefont {Lettow}, \citenamefont {Pfab}, \citenamefont {Renn},
  \citenamefont {Ikonen}, \citenamefont {Gotzinger},\ and\ \citenamefont
  {Sandoghdar}}]{ahtee_molecules_2009}%
  \BibitemOpen
  \bibfield  {author} {\bibinfo {author} {\bibfnamefont {V.}~\bibnamefont
  {Ahtee}}, \bibinfo {author} {\bibfnamefont {R.}~\bibnamefont {Lettow}},
  \bibinfo {author} {\bibfnamefont {R.}~\bibnamefont {Pfab}}, \bibinfo {author}
  {\bibfnamefont {A.}~\bibnamefont {Renn}}, \bibinfo {author} {\bibfnamefont
  {E.}~\bibnamefont {Ikonen}}, \bibinfo {author} {\bibfnamefont
  {S.}~\bibnamefont {Gotzinger}}, \ and\ \bibinfo {author} {\bibfnamefont
  {V.}~\bibnamefont {Sandoghdar}},\ }\bibfield  {title} {\enquote {\bibinfo
  {title} {Molecules as sources for indistinguishable single photons},}\ }\href
  {http://www.tandfonline.com/doi/pdf/10.1080/09500340802464657} {\bibfield
  {journal} {\bibinfo  {journal} {J. Mod. Opt.}\ }\textbf {\bibinfo {volume}
  {56}},\ \bibinfo {pages} {161--166} (\bibinfo {year} {2009})}\BibitemShut
  {NoStop}%
\bibitem [{\citenamefont {Polisseni}\ \emph {et~al.}(2016)\citenamefont
  {Polisseni}, \citenamefont {Major}, \citenamefont {Boissier}, \citenamefont
  {Grandi}, \citenamefont {Clark},\ and\ \citenamefont
  {Hinds}}]{polisseni_stable_2016}%
  \BibitemOpen
  \bibfield  {author} {\bibinfo {author} {\bibfnamefont {C.}~\bibnamefont
  {Polisseni}}, \bibinfo {author} {\bibfnamefont {K.~D.}\ \bibnamefont
  {Major}}, \bibinfo {author} {\bibfnamefont {S.}~\bibnamefont {Boissier}},
  \bibinfo {author} {\bibfnamefont {S.}~\bibnamefont {Grandi}}, \bibinfo
  {author} {\bibfnamefont {A.~S.}\ \bibnamefont {Clark}}, \ and\ \bibinfo
  {author} {\bibfnamefont {E.~A.}\ \bibnamefont {Hinds}},\ }\bibfield  {title}
  {{\selectlanguage {english}\enquote {\bibinfo {title} {Stable, single-photon
  emitter in a thin organic crystal for application to quantum-photonic
  devices},}\ }}\href {\doibase 10.1364/OE.24.005615} {\bibfield  {journal}
  {\bibinfo  {journal} {Opt. Express}\ }\textbf {\bibinfo {volume} {24}},\
  \bibinfo {pages} {5615} (\bibinfo {year} {2016})}\BibitemShut {NoStop}%
\bibitem [{\citenamefont {Witthaut}\ \emph {et~al.}(2012)\citenamefont
  {Witthaut}, \citenamefont {Lukin},\ and\ \citenamefont
  {S{\o}rensen}}]{witthaut_photon_2012}%
  \BibitemOpen
  \bibfield  {author} {\bibinfo {author} {\bibfnamefont {D.}~\bibnamefont
  {Witthaut}}, \bibinfo {author} {\bibfnamefont {M.~D.}\ \bibnamefont {Lukin}},
  \ and\ \bibinfo {author} {\bibfnamefont {A.~S.}\ \bibnamefont
  {S{\o}rensen}},\ }\bibfield  {title} {{\selectlanguage {english}\enquote
  {\bibinfo {title} {Photon sorters and {QND} detectors using single photon
  emitters},}\ }}\href {\doibase 10.1209/0295-5075/97/50007} {\bibfield
  {journal} {\bibinfo  {journal} {EPL (Europhysics Letters)}\ }\textbf
  {\bibinfo {volume} {97}},\ \bibinfo {pages} {50007} (\bibinfo {year}
  {2012})}\BibitemShut {NoStop}%
\bibitem [{\citenamefont {Ralph}\ \emph {et~al.}(2015)\citenamefont {Ralph},
  \citenamefont {S\"{o}llner}, \citenamefont {Mahmoodian}, \citenamefont
  {White},\ and\ \citenamefont {Lodahl}}]{ralph_photon_2015}%
  \BibitemOpen
  \bibfield  {author} {\bibinfo {author} {\bibfnamefont {T. C.}\ \bibnamefont
  {Ralph}}, \bibinfo {author} {\bibfnamefont {I.}~\bibnamefont {S\"{o}llner}},
  \bibinfo {author} {\bibfnamefont {S.}~\bibnamefont {Mahmoodian}}, \bibinfo
  {author} {\bibfnamefont {A. G.}\ \bibnamefont {White}}, \ and\ \bibinfo
  {author} {\bibfnamefont {P.}~\bibnamefont {Lodahl}},\ }\bibfield  {title}
  {\enquote {\bibinfo {title} {Photon {Sorting}, {Efficient} {Bell}
  {Measurements}, and a {Deterministic} {Controlled}-${Z}$ {Gate} {Using} a
  {Passive} {Two}-{Level} {Nonlinearity}},}\ }\href {\doibase
  10.1103/PhysRevLett.114.173603} {\bibfield  {journal} {\bibinfo  {journal}
  {Phys. Rev. Lett.}\ }\textbf {\bibinfo {volume} {114}},\ \bibinfo {pages}
  {173603} (\bibinfo {year} {2015})}\BibitemShut {NoStop}%
\bibitem [{\citenamefont {Hegerfeldt}\ and\ \citenamefont
  {Seidel}(2003)}]{hegerfeldt_blinking_2003}%
  \BibitemOpen
  \bibfield  {author} {\bibinfo {author} {\bibfnamefont {Gerhard~C.}\
  \bibnamefont {Hegerfeldt}}\ and\ \bibinfo {author} {\bibfnamefont {Dirk}\
  \bibnamefont {Seidel}},\ }\bibfield  {title} {\enquote {\bibinfo {title}
  {Blinking molecules: {Determination} of photophysical parameters from the
  intensity correlation function},}\ }\href {\doibase 10.1063/1.1563615}
  {\bibfield  {journal} {\bibinfo  {journal} {J. Chem. Phys.}\ }\textbf
  {\bibinfo {volume} {118}},\ \bibinfo {pages} {7741--7746} (\bibinfo {year}
  {2003})}\BibitemShut {NoStop}%
\bibitem [{\citenamefont {Bransden}\ and\ \citenamefont
  {Joachain}(2003)}]{book_bransden}%
  \BibitemOpen
  \bibfield  {author} {\bibinfo {author} {\bibfnamefont {B.~H.}\ \bibnamefont
  {Bransden}}\ and\ \bibinfo {author} {\bibfnamefont {C.~J.}\ \bibnamefont
  {Joachain}},\ }\href@noop {} {\emph {\bibinfo {title} {Physics of Atoms and
  Molecules}}}\ (\bibinfo  {publisher} {Prentice Hall},\ \bibinfo {year}
  {2003})\BibitemShut {NoStop}%
\end{thebibliography}%

\end{document}